\documentclass[useAMS,usenatbib]{mn2e}
\usepackage{graphicx}
\usepackage{epstopdf}

\newcommand{\spose}[1]{\hbox to 0pt{#1\hss}}
\newcommand{\approxpropto}{\mathrel{\spose{\lower 3pt\hbox{$\sim$}}
	\raise 2.0pt\hbox{$\propto$}}}
\def\approxgt{\mathrel{\spose{\lower 3pt\hbox{$\sim$}}
	\raise 2.0pt\hbox{$>$}}}
\def\approxlt{\mathrel{\spose{\lower 3pt\hbox{$\sim$}}
	\raise 2.0pt\hbox{$<$}}}

\def \hkpc{\,h^{-1} \hbox{kpc}}
\def \hMpc{\,h^{-1} \hbox{Mpc}}
\def \YSZunits{\,h^{-2} \, \hbox{Mpc}^{2}}
\def \YXunits{\,h^{-1} \, \hbox{M}_{\odot} \, \hbox{keV}} 
\def \hMsol{\,h^{-1} \hbox{M}_{\odot}}

\def \h2gcm3{h^{2} \, \hbox{g}\,\hbox{cm}^{-3}}

\def \kevcm2{\, \hbox{keV} \, \hbox{cm}^{2}}
\def \cm3{\, \hbox{cm}^{-3}}
\def \keVcm2{\, \hbox{keVcm}^{2}}

 \title[SZ clusters in Millennium Gas simulations]
       {Sunyaev--Zel'dovich clusters in Millennium Gas simulations}

\author[S.T.Kay et al.]
            {Scott T. Kay,$^{1}$\thanks{E-mail: Scott.Kay@manchester.ac.uk}
	     Michael W. Peel,$^{1}$ C.~J. Short,$^{2}$ Peter A. Thomas,$^{2}$ \newauthor
	     Owain E. Young,$^{2}$ Richard A. Battye,$^{1}$ Andrew R. Liddle$^{2}$ and Frazer R. Pearce$^{3}$\\
	$^{1}$Jodrell Bank Centre for Astrophysics, School of Physics and Astronomy, The University of Manchester, Manchester M13 9PL\\
	$^{2}$Astronomy Centre, Department of Physics and Astronomy, University of Sussex, Brighton BN1 9QH\\ 
	$^{3}$Department of Physics and Astronomy, University of Nottingham, Nottingham NG7 2RD
	}

\begin{document}
\date{This draft was generated on \today}

\pagerange{\pageref{firstpage}--\pageref{lastpage}} \pubyear{2005}

\maketitle

\label{firstpage}
\begin{abstract}
Large surveys using the Sunyaev-Zel'dovich (SZ) effect to find clusters 
of galaxies are now starting to yield large numbers of systems out to 
high redshift, many of which are new discoveries. In order to provide 
theoretical interpretation for the release of the full SZ cluster samples
over the next few years, we have exploited the large-volume Millennium 
Gas cosmological $N$-body hydrodynamics simulations to study the SZ cluster
population at low and high redshift, for three models with varying gas
physics. We confirm previous results using smaller samples that the 
intrinsic (spherical) $Y_{500}-M_{500}$ relation has very little scatter 
($\sigma_{\log_{10}Y}\simeq 0.04$), is insensitive to cluster gas physics and 
evolves to redshift one in accord with self-similar expectations. Our 
{\it pre-heating} and {\it feedback} models predict scaling relations 
that are in excellent agreement with the recent analysis from combined {\it Planck}
and {\it XMM-Newton} data by the Planck Collaboration. 
This agreement is largely preserved when $r_{500}$ and $M_{500}$ are
derived using the hydrostatic mass proxy, $Y_{\rm X,500}$, albeit with significantly
reduced scatter ($\sigma_{\log_{10}Y}\simeq 0.02$), a result that is due to the
tight correlation between $Y_{500}$ and $Y_{\rm X,500}$. Interestingly, this 
assumption also hides any bias in the relation due to dynamical activity. We 
also assess the importance of projection effects from large-scale structure along
the line-of-sight, by extracting cluster  $Y_{500}$ values from fifty simulated 
$5^{\circ} \times 5^{\circ}$ 
sky  maps. Once the (model-dependent) mean signal is subtracted from the maps 
we find that the integrated SZ signal is unbiased with respect to the underlying clusters,
although the scatter in the (cylindrical) $Y_{500}-M_{500}$ relation increases in the 
{\it pre-heating} case, where a significant amount of energy was injected into the 
intergalactic medium at high redshift. Finally, we study the hot gas pressure profiles
to investigate the origin of the SZ signal and find that the largest contribution comes
from radii close to $r_{500}$ in all cases. The profiles themselves are well described by generalised
Navarro, Frenk \& White profiles but there is significant cluster-to-cluster scatter.
In conclusion, our results support the notion that $Y_{500}$ is a robust mass proxy for
use in cosmological analyses with clusters.
\end{abstract}


\section{Introduction}

The Sunyaev-Zel'dovich (SZ) effect \citep{Sunyaev72} is a powerful method
for discovering new clusters of galaxies. It arises generically due to the 
scattering of Cosmic Microwave Background (CMB) photons off free electrons,
leading to a predictable spectral distortion in the CMB that is, in the
non-relativistic limit, linearly dependent on the line integral of the electron pressure 
\citep{Birkinshaw99}.
In modern theories of structure formation, the dominant contribution to the
SZ signal comes from the intracluster medium (ICM), a diffuse plasma within
clusters that is approximately in hydrostatic equilibrium within the dark-matter
dominated potential (see \citealt{Voit05,Allen11} for 
recent reviews). The key SZ observable is the $Y$ parameter, defined as 
\begin{equation}
Y = \int \, y \, {\rm d}\Omega,
\end{equation}
where the integral is performed over the solid angle subtended by the cluster.
The Compton-$y$ parameter is determined by the thermal structure of the ICM
\begin{equation}
y = {\sigma_{\rm T} k \over m_{\rm e} c^2} \, \int \, n_{\rm e}T_{\rm e} \, {\rm d}l,
\end{equation}
where $n_{\rm e}$ and $T_{\rm e}$ are the density and temperature of the free
electrons respectively and ${\rm d}l$ is the differential line element along
the line-of-sight.
Since $Y$ can be expressed as a volume integral of the pressure (when the
redshift and cosmological parameters are specified),
it measures the total thermal energy of the gas, a property that ought to be
strongly correlated with the cluster's mass through the virial theorem. This 
means that $Y$ ought to be relatively insensitive to the complex micro-physics
taking place in the cluster core, unlike other global properties such 
as X-ray luminosity. 

% OBSERVATIONS

Early observational studies confirmed the detection of an SZ 
signal towards known massive clusters of galaxies and used this to
estimate the Hubble constant (e.g. \citealt{Jones93,Birkinshaw94}).
Over the past decade, SZ observations of known bright X-ray bright clusters have
become routine, allowing the investigation of cluster scaling relations to be
performed
(e.g. \citealt{McCarthy03b,Benson04,Morandi07,Bonamente08,Huang10,Shimwell11,Lancaster11}). 
One potential shortcoming of this approach is that the samples are X-ray selected 
and therefore biased towards luminous, cool-core systems at low redshift.

In the past few years, SZ science has entered the exciting new phase of blind
surveys, where detections of new clusters have become possible \citep{Staniszewski09}.
Indeed, SZ surveys are now yielding large numbers of SZ-selected clusters, 
many of them new detections, especially from 
the South Pole Telescope (SPT; \citealt{Vanderlinde10,Andersson11,Williamson11}),
the Atacama Cosmology Telescope (ACT; \citealt{Marriage11,Sehgal11}) 
and the {\it Planck} satellite \citep{Planck11a,Planck11b,Planck11c}.
Since the SZ effect is effectively independent of redshift, the SZ selection
function tends to favour higher redshift systems than the X-ray counterpart,
assuming similar angular resolution. As a result, the new blind SZ surveys are starting to 
find new massive systems at $z \sim 1$ \citep{Planck11d,Foley11,Menanteau11}. 
In the near future, we should expect to see 
these numbers increase substantially as the full survey results are published, 
nicely complementing X-ray surveys such as the MAssive Cluster Survey
(MACS; \citealt{Ebeling01}) and the {\it XMM} Cluster Survey 
(XCS; \citealt{Romer01,Mehrtens11}). 
Such complementarity will be further exploited with the next generation of 
X-ray surveys (e.g. with eROSITA) and millimetric telescopes (e.g. CCAT; 
see \citealt{Golwala09}).

% MODELS

One of the main goals of SZ surveys is to measure cosmological parameters
(e.g. \citealt{Barbosa96,Carlstrom02,Battye03}). 
Central to the cosmological application of SZ surveys is the scaling
relation between the observables ($Y$ and redshift, $z$) and cluster 
mass, $M$. Under the assumption that
clusters form a self-similar population \citep{Kaiser86} the SZ flux
should scale as $Y \propto M^{5/3} H(z)^{2/3}$, when measured within a
radius enclosing a mean density that is a constant multiple of the critical
density of the Universe. Early theoretical studies combined such 
simple scaling relations with the Press-Schechter formalism \citep{Press74}
to predict the SZ evolution of the cluster population in a variety of 
cosmological models
(e.g. \citealt{Cole88,Bartlett94,Barbosa96,Eke96,Aghanim97,Kay01,Battye03}). 
More recently, attention has turned to more detailed studies of how cluster
gas physics impacts upon SZ scaling relations, both using semi-analytic
models (e.g. \citealt{McCarthy03a,McCarthy03b,Shaw08}) 
and full cosmological $N$-body/hydrodynamic simulations
\citep{daSilva00,White02,daSilva04,Motl05,Nagai06,Bonaldi07,Hallman07,Aghanim09,Battaglia11}.
Simulations are now also being used to investigate the effects of mergers
on SZ scaling relations \citep{Poole07,Wik08,Yang10,Krause12}. A generic result
from these studies is that the self-similar description appears to be 
approximately valid on cluster scales ($M>10^{14}\hMsol$)
but in detail, differences are seen between the models that are
due to the effects of non-gravitational physics (cooling and heating processes), 
especially at low mass where the gas fraction is depleted.

% THIS WORK

Two of the main shortcomings in previous simulation studies are the relatively 
small samples (that are sometimes restricted to lower-mass clusters) and a limited
range of (uncertain) cluster gas physics models, often not calibrated to match
X-ray data. Some studies may satisfy one of these criteria but usually not both. 
A new generation of simulations are now starting to overcome both shortcomings. 
\citet{Stanek10} recently presented results from two of the {\it Millennium Gas} 
simulations \citep{Hartley08}, large-volume runs based on the Millennium Simulation 
\citep{Springel05a} with varying gas physics.  
These simulations are sufficiently large to enable the full range of cluster
masses ($10^{14}-10^{15} \hMsol$) to be studied and one of the runs, where the gas
was pre-heated at high redshift, is able to match the mean X-ray luminosity-temperature
relation at $z=0$ \citep{Hartley08}. Although the work of \citet{Stanek10} was focused
on the more general issue of multi-variate scaling relations, they presented
results for the SZ $Y-M$ relation measured within a radius corresponding to 
a mean internal density equal to 200 times the critical density, $r_{200}$.

The aim of this paper is to use these Millennium Gas simulations to focus in
more detail on predictions of the SZ effect and, in particular, the 
$Y-M$ relation for clusters. Our paper builds on the \citet{Stanek10} work
in three important ways. Firstly, we add a third model that includes a more
realistic treatment of feedback, both from supernovae and active galactic nuclei. 
This model has already been shown to successfully match many of the X-ray properties of
non-cool core clusters \citep{Short09,Short10}. Secondly, we include in our analysis 
simulated maps of the full SZ effect along the line-of-sight, to assess the
projection effects of large-scale structure. Finally, we attempt to produce results for our
$Y-M$ scaling relations using methods that are more closely matched with observations.
In particular, we present our results for the smaller $r_{500}$ and investigate the
impact of assuming hydrostatic equilibrium and a mass proxy ($Y_{\rm X}$, 
\citealt{Kravtsov06}) on the $Y-M$ relation.

% LAYOUT

We organise the remainder of the paper as follows. 
In Section~\ref{sec:simdetails} we outline the simulation details and
our methods used to define cluster properties. We also present some basic
properties of the sample and SZ maps. 
Sections~\ref{sec:pressureprofiles}, \ref{sec:scalingrelations} and \ref{sec:hse}
contain our main results: in Section~\ref{sec:pressureprofiles} 
we present an analysis of the hot gas pressure profiles, before going on
to study SZ scaling relations in Section~\ref{sec:scalingrelations} and the 
impact of hydrostatic bias in Section~\ref{sec:hse}. 
%In the latter case, we start by presenting the true scaling relations then
%investigate the effects of hydrostatic mass estimates and the use of 
%$Y_{\rm X}$ as a mass proxy. We also compare our
%work to results from both observations and previous simulations, before studying
%projection effects along the line-of-sight.
Finally, in Section~\ref{sec:summary} we summarise our main conclusions and
outline future work.
%\vspace{-0.5cm}

\section{Simulation Details}
\label{sec:simdetails}

Our results are drawn from the Millennium Gas simulations, a set of
large, cosmological hydrodynamics simulations of the $\Lambda$CDM 
cosmology ($\Omega_{\rm m}=0.25$, $\Omega_{\Lambda}=0.75$, 
$\Omega_{\rm b}=0.045$, $h=0.73$, $\sigma_8=0.9$). In this section
we summarise the details of these simulations and present our methods
for constructing simulated cluster properties and SZ sky maps. 

\subsection{Millennium Gas simulations}

The Millennium Gas simulations (hereafter MGS; see
\citealt{Hartley08,Stanek09,Stanek10,Short10,Young11}) 
were constructed to provide hydrodynamic
versions of the Virgo Consortium's dark matter Millennium Simulation 
(hereafter MS; \citealt{Springel05a}). 
The simulations were therefore started from the same 
realisation of the large-scale density field within the same comoving box-size,
$L=500 \hMpc$ and used the same set of cosmological parameters.  
The MGS were run with the publicly-available {\sc gadget2}
$N$-body/hydrodynamics code \citep{Springel05b}. Due to the increased computational
requirements from the inclusion of gas particles, the simulations were run with fewer
($5 \times 10^8$ each of gas and dark matter) particles in total than the MS. 
The particle masses were therefore set to $m_{\rm gas}=3.1\times 10^{9} \hMsol$
and $m_{\rm dm}=1.4\times 10^{10} \hMsol$ for the gas and dark matter
respectively. Gravitational forces were softened at small separations using an equivalent
Plummer softening length of $\epsilon = 100 \hkpc$, held fixed in comoving co-ordinates. At
low redshift ($z<3$) the softening was then fixed to $\epsilon = 25 \hkpc$ in physical
co-ordinates.

Two versions of the MGS were run with the above properties. Both runs started from
identical initial conditions but differed in the way the gas was evolved. In the first
run, the gas was modelled as an ideal non-radiative fluid. In addition to gravitational
forces, the gas could undergo adiabatic changes in regions of non-zero pressure gradients,
modelled using the Smoothed Particle Hydrodynamics formalism (SPH; see \citealt{Springel02}
for the version of SPH used in {\sc gadget2}). Additionally, in regions where the flow was 
convergent the bulk kinetic energy of the gas is converted into internal energy using an
artificial viscosity term; this is essential to capture shocks and thus generate 
quasi-hydrostatic atmospheres within virialised dark matter haloes. In accord with previous
studies (e.g. \citealt{Short10}) we refer to this simulation as the GO 
({\it Gravitation Only}) model.

It is well known that a non-radiative description of intracluster gas does not agree with
the observed X-ray properties of clusters, especially at low masses, where an excess
of core entropy is required to produce a steeper X-ray luminosity-temperature relation 
(e.g. \citealt{Voit05}). A simple method capable of generating this excess entropy
is to {\it pre-heat} the gas at high redshift before cluster collapse 
\citep{Kaiser91,Evrard91}. We implemented this method in a second simulation by
raising the minimum entropy 
\footnote{In the usual way, we take entropy to mean the quantity $K=kT/n_{\rm e}^{2/3}$,
where $T$ is the gas temperature and $n_{\rm e}$ the free electron density.}
of the gas (by increasing its temperature) to $K_{\rm min}=200\,{\rm keV} \, {\rm cm}^2$ at $z=4$.
The entropy level was chosen so as to match the mean $z=0$ X-ray 
luminosity-temperature relation \citep{Hartley08}. We also included radiative 
cooling, an entropy sink. However, this made very little difference,
as the cooling time of the pre-heated gas is very long compared to the Hubble time and 
therefore gas could  no longer cool and form stars before the end of the simulation. 
We refer to this simulation using the label PC, for {\it Pre-heating plus Cooling}.

We also consider a third model when analysing the SZ properties from individual
clusters. This is the {\it Feedback Only} (hereafter FO) model developed by 
\citet{Short09} and then applied to MGS clusters by \citet{Short10}, where full
details of the method may be found. Briefly, it uses the semi-analytic galaxy 
formation model of \citet{DeLucia07}, run on dark-matter-only resimulations of
MS clusters, to provide information on the effects of star formation and feedback on the 
intracluster gas. The model works as follows. Galaxy merger trees are first
generated by applying the semi-analytic model to the dark matter distribution. 
Various properties of the galaxies (such as their position, stellar mass and
black hole mass) are stored at each snapshot of the simulation. The
clusters are then re-simulated with gas, assuming that the gas particles have
zero gravitational mass; this guarantees that the dark matter distributions (and
therefore galaxy positions) are identical to those in the parent 
dark-matter-only simulation. At
each snapshot time, two important changes are made to the gas. Firstly the increase
in stellar mass of each cluster galaxy is used to convert local intracluster gas
into stars, a requirement for generating sensible stellar and gas fractions
\citep{Young11}. This change in stellar mass is also used to heat the gas
from supernova explosions. Secondly, any increase in black hole mass is 
used to heat the gas on the basis that such accretion leads to an Active Galactic Nucleus 
(AGN). The heating rate, known as AGN feedback, is taken from \citet{Bower08}
and is given by
\begin{equation}
\dot{E}_{\rm feed} = {\rm min} \left( \epsilon_{\rm SMBH}L_{\rm Edd}, 
\epsilon_{\rm r}\dot{M}_{\rm BH} c^2 \right),
\end{equation}
where $\epsilon_{\rm SMBH}=0.02$ dictates the maximum heating rate
(in units of the Eddington luminosity)
and $\epsilon_{\rm r}=0.1$ is the efficiency with which the accreted 
mass is converted into feedback energy. This is particularly 
important because AGN
are the dominant feedback mechanism on cluster scales.

We analyse the same sample of 337 clusters studied by \citet{Short10},
comprising all objects in the MS with virial mass 
$M_{\rm vir}>5\times 10^{14}\hMsol$ and a random sample at lower mass
($1.7\times 10^{13}\hMsol \leq M_{\rm vir} \leq 5\times 10^{14}\hMsol$)
chosen such that there were a fixed number of objects within each logarithmic
mass bin. The FO model successfully generates the required excess entropy
of the low redshift population and provides a good match to the structural
properties of non-cool core clusters. The main shortcoming of this model is
that it neglects the effects of radiative cooling and therefore cannot 
reproduce the most X-ray luminous cool core population \citep{Short10}.
This failure may not be as serious as it seems, however, since there is
some evidence that the X-ray cool core population diminishes with increasing redshift,
both from observations (e.g. \citealt{Maughan11}) and simulations (e.g. \citealt{Kay07}).
Furthermore, as we will demonstrate, the SZ $Y$ parameter (which measures the global thermal 
energy of the intracluster gas) is reasonably insensitive to changes in gas physics
that predominantly affect the cluster core. Issues relevant to this study where 
cooling could impact upon our results are the degree to which the ICM is hydrostatic 
and the effect of gas clumping on the X-ray quantity, $Y_{\rm X}$, used as a cluster mass proxy. 
We note that a first step towards including radiative cooling in the model has been 
made and shows promising results \citep{Short12}. Ultimately, a fully self-consistent scheme is desirable, 
where the same cooling and heating rates are used in both the semi-analytic model and 
hydrodynamic simulation.

\subsection{Cluster definitions and estimation of global properties}

Clusters are defined in exactly the same way as in \citet{Kay07}.
Firstly, a friends-of-friends code is run on the dark matter particles for each
snapshot. The dimensionless linking length (in units of the mean inter-particle
separation) is set to $b=0.1$, chosen to minimise the probability of linking 
two haloes together outside of their respective virial radii. The dark matter 
particle with the most negative gravitational potential energy is then identified 
for each group and this is taken to be the centre. 

In the next stage, a sphere is centred on each friends-of-friends group and its
radius increased until the total mass (from dark matter, gas and stars, when present)
satisfies
\begin{equation}
M_{\Delta}={4 \pi \over 3} \, r_{\Delta}^3 \, \Delta \, \rho_{\rm cr}(z),
\end{equation}
where $r_{\Delta}$ is the proper radius of the sphere, $\Delta$ is a specified
density contrast, $\rho_{\rm cr}(z)=(3H_0^2/8\pi G)E(z)^2$ is the critical density and  
$E(z)^2=\Omega_{\rm m}(1+z)^3+\Omega_{\Lambda}$ for a flat universe. 
We assume $\Delta=500$ for the main results in this study as this value is commonly 
adopted for observational studies (some of which we will compare to) because $r_{500}$ is 
sufficiently large to make many integrated properties insensitive to variations in core 
structure, while also being small enough to be within reach for detailed X-ray 
observations of many objects. We occasionally use the value of $\Delta$ 
appropriate for the virial radius, $r_{\rm vir}$, as defined by the 
spherical top-hat collapse model. This is a redshift-dependent quantity, 
$\Delta=\Delta_{\rm c}(z)$, which we calculate using the fitting formula
given by \citet{Bryan98}. Note that at $z=0$, $\Delta_{\rm c} \simeq 94$ and 
$r_{\rm vir}\simeq 2\, r_{500}$.
 
Once the cluster's mass and radius is defined, we calculate various properties of the
hot gas, the most important being the SZ flux. The frequency independent part is given
by
\begin{equation}
Y_{500} = {1 \over D_{\rm A}^2} \, {\sigma_{\rm T} \over m_{\rm e}c^2} \,
\int n_{\rm e}kT_{\rm e} \, {\rm d}V,
\label{eqn:szY}
\end{equation}
where $D_{\rm A}$ is the (cosmology-dependent) angular diameter
distance to the cluster and the integral is performed over the entire 
cluster sphere. To simplify matters, we re-define the integrated SZ $Y$ parameter 
\begin{equation}
Y_{500} \, D_{\rm A}^2 \rightarrow Y_{500},
\end{equation}
since this combination is directly proportional to the integrated thermal energy of the gas
which is the physical property of interest. Note
that the dimensions of $Y_{500}$ are now that of area; we will therefore present
values in $\YSZunits$ units. The value of $Y_{500}$ is estimated for each
cluster using
\begin{equation}
Y_{500}
= \left( 
{\sigma_{\rm T} k m_{\rm gas} \over \mu_{\rm e} m_{\rm H} m_{\rm e} c^2 }
\right) \, \sum_{i=1}^{N_{\rm hot}} \, T_i,
\end{equation}
where the sum runs over all hot ($T>10^5$K) gas particles within $r_{500}$,
with mass $m_{\rm gas}$ and temperature $T_i$. We adopt the value $\mu_{\rm e}=1.14$
for the mean molecular weight per free electron, appropriate for a fully ionised plasma of
hydrogen (with mass fraction $X=0.76$) and helium (with mass fraction $Y=1-X$). We also
assume equipartition of energy between the electrons and nuclei, thus $T=T_{\rm e}$.

We estimate the X-ray temperature of the ICM using the spectroscopic-like temperature $T_{\rm sl}$
\citep{Mazzotta04}, appropriate for bremsstrahlung in hot ($kT>3 \, {\rm keV}$) clusters
\begin{equation}
T_{\rm sl} = { \sum_{i=1}^{N_{\rm hot}} \,\rho_i T_i^{1/4} 
\over \sum_{i=1}^{N_{\rm hot}} \,\rho_i T_i^{-3/4}},
\label{eqn:tsl}
\end{equation}
where $\rho_i$ is the density of particle $i$ and in this case 
the sum runs over all
hot gas particles with $kT_i>0.5 $keV.
We measure $T_{\rm sl}$ in the region outside the cluster core ($x_{\rm core}<r/r_{500}<1$, where
$x_{\rm core}=0.1$ for the GO and PC models, and 0.15 for the FO model
\footnote{
The GO/PC and FO data were processed independently and different choices for 
$x_{\rm core}$ were made at those times. However, the effect of this difference on $T_{\rm sl}$ 
is small; we checked by re-calculating the GO/PC temperatures at $z=0$ using $x_{\rm core}=0.15$ 
and found only a 2-3 per cent increase, on average.
}
)
to provide a closer match to observed X-ray temperature measurements (where a larger variation
in core temperature is seen than in our simulations).

A quantity related to $Y_{500}$ is $Y_{\rm X,500} \propto M_{\rm gas}T_{\rm X}$, 
estimated from X-ray data. Introduced by \citet{Kravtsov06}, it was shown to be a low-scatter 
proxy for cluster mass (due to scatter in X-ray temperature being negatively correlated with 
scatter in gas mass). We estimate this quantity as
\begin{equation}
Y_{\rm X,500} = \left( 
{\sigma_{\rm T} k \over \mu_{\rm e} m_{\rm H} m_{\rm e} c^2 }
\right) \, M_{\rm gas,500} \, T_{\rm sl},
\end{equation}
where $M_{\rm gas,500}$ is the mass of hot gas within $r_{500}$, although
we occasionally present $Y_{\rm X,500}$ in its native ($h^{-1} \, {\rm M}_{\odot} \,{\rm keV}$)
units, i.e. simply assuming $Y_{\rm X,500}=M_{\rm gas,500}kT_{\rm sl}$.
The main difference between $Y$ and $Y_{\rm X}$ is that the former depends
on the mass-weighted temperature while the latter depends on the X-ray temperature,
which is more heavily weighted by lower entropy gas \citep{Mazzotta04}. Comparing
$Y$ with $Y_{\rm X}$ therefore implicitly tests the {\it clumpiness} of the ICM
since clumpy gas will be cooler and therefore lower the X-ray temperature relative
to the mass-weighted temperature (e.g. \citealt{Kay08}). As we show below, this effect
is model dependent but is of minimal importance in the PC and FO simulations.

\subsection{Cluster sample}

\begin{table}
\caption{Number of clusters in our samples at redshifts, $z=0,0.5$ and $1$. 
Column 1 gives the model label and column 2 the redshift. Column 3
lists the total number of clusters in each sample with $M_{500}>10^{14}\hMsol$,
while columns 4 and 5 sub-divide the sample into the {\it regular} and 
{\it disturbed} populations respectively, using the $s$ parameter defined
in equation~(\ref{eqn:substat}).}
\begin{center}
\begin{tabular}{lcccc}
\hline
Model & Redshift & $N_{\rm clus}$ & $N_{\rm clus}(s \leq 0.1)$ & $N_{\rm clus}(s>0.1)$\\
\hline
GO & 0.0 &           1110 & 986 & 124\\
   & 0.5 & \phantom{0}567 & 457 & 110\\
   & 1.0 & \phantom{0}139 & 103 & \phantom{0}36\\
\hline
PC & 0.0 & \phantom{0}883 & 799 & \phantom{0}84\\
   & 0.5 & \phantom{0}436 & 355 & \phantom{0}81\\
   & 1.0 & \phantom{0}102 & \phantom{0}78 & \phantom{0}24\\
\hline
FO & 0.0 & \phantom{0}188 & 154 & \phantom{0}34\\
   & 0.5 & \phantom{0}148 & 122 & \phantom{0}26\\
   & 1.0 & \phantom{00}75 & \phantom{0}51 & \phantom{0}24\\
\hline
\end{tabular}
\end{center}
\label{tab:sample}
\end{table}

Table~\ref{tab:sample} summarises the number of clusters in each of the runs
at redshifts, $z=0,0.5$ and $1$. For our fiducial sample we have employed a lower mass
cut of $M_{500} > 10^{14} \hMsol$, a useful limit for comparing with SZ cluster data.
The GO and PC simulations have similar numbers, although the latter is slightly smaller 
due to the effect of pre-heating on the gas fraction \citep{Stanek09}. Note the number of 
clusters at $z=1$ is around an order of magnitude lower than at $z=0$. There are 
significantly fewer FO clusters at any given redshift due to the fact that it is not
a volume-limited sample. The drop in number at high redshift is 
not as severe in this case as the mean mass of the sample is higher and so a smaller
fraction of clusters drops below the imposed mass limit.

We also consider the effect of ongoing mergers by splitting our sample into 
{\it regular} and {\it disturbed}  sub-samples, using a simple estimator
known as the {\it substructure statistic}
\citep{Thomas98,Kay07}, defined as
\begin{equation}
s = { | \mathbf{r}_{\phi} - \mathbf{r}_{\rm cm} | \over r_{500}},
\label{eqn:substat}
\end{equation}
where $\mathbf{r}_{\phi}$ is the position of the cluster centre 
(defined here to be the position of the dark matter particle with the most
negative potential, $\phi$) and $\mathbf{r}_{\rm cm}$ is the 
centre-of-mass. We define those clusters with $s>0.1$
as disturbed systems and those with $s\leq 0.1$ as regular systems,
although note that this terminology is strictly for convenience as all clusters
are disturbed to some degree. In practice, this value delineates those that
are clearly undergoing significant mergers, as discussed in \citet{Kay07}. The fraction of
disturbed clusters increases with redshift in all models, from around 10 per cent at $z=0$
to 25 per cent at $z=1$, in the GO and PC models. Again, the different method for
cluster selection in the FO model modifies the result but nevertheless the trend 
of increasing disturbed fraction with redshift is still seen.

\subsection{Cluster profiles}
\label{sec:clusprofiles}

We discuss hot gas pressure profiles in Section~\ref{sec:pressureprofiles} as
these are important for understanding the relative contribution to the SZ signal from 
different radii. The profiles are constructed by first identifying all hot gas 
particles within a radius $r_{500}$ of the cluster centre. This sphere is then 
sub-divided into spherical shells with fixed radial thickness in $\log_{10}(x)$, where 
$x=r/r_{500}$. The pressure within the shell is then estimated using a mass-weighted
average 
\begin{equation}
P(x) = {1 \over V(x)} {k \over \mu m_{\rm H}} \, \sum_{i=1}^{N_{\rm shell}} \, m_i T_i,
\label{eqn:pprof}
\end{equation}
where the sum runs over all hot gas particles within the shell at radial position 
$x$, $V$ is the volume of the shell and $\mu=0.59$ is the mean molecular weight for 
an ionised plasma (assuming zero metallicity).

\subsection{Cluster maps}
\label{sec:clustermaps}

\begin{figure*}
\centering
\includegraphics[width=18cm]{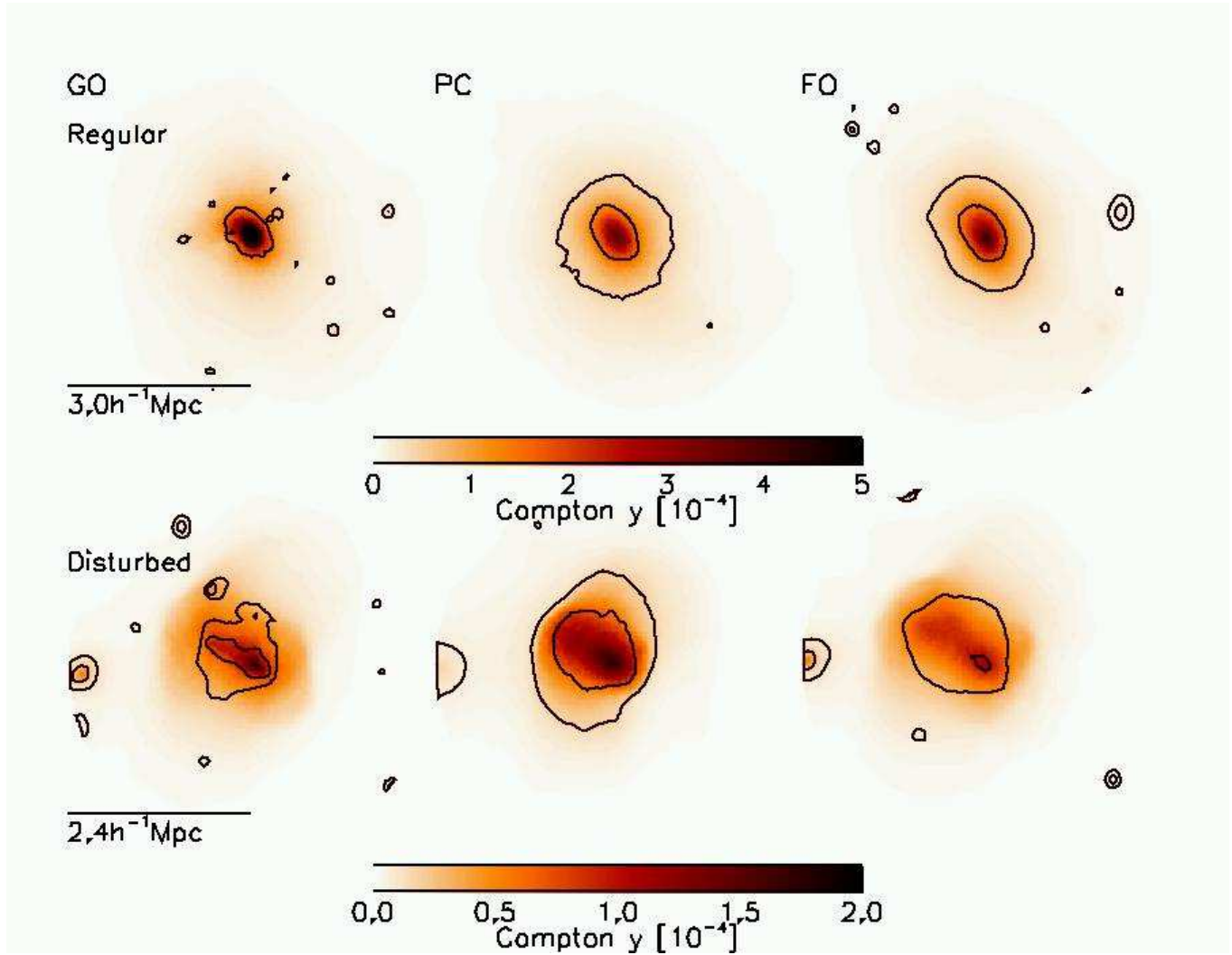}
\caption{Top panels: Compton-$y$ maps for the most massive cluster 
at  $z=0$ in the MS ($M_{\rm vir} \simeq 2.9 \times 10^{15} \hMsol$, which we classify 
as regular) with X-ray surface brightness contours overlaid. Results are shown, 
from left to right, for the cluster in the GO, PC and FO models respectively. 
Images in the bottom panels are similar except a massive disturbed cluster
(with $M_{\rm vir} \simeq 1.5 \times 10^{15} \hMsol$) is shown. 
Each panel is $2r_{\rm vir}$ across and the scale and value of $r_{\rm vir}$
is shown in the left-hand panels. The range of
$y$ values is given for each cluster in the scale at the bottom of each row; note the disturbed
cluster has a lower maximum value than the regular cluster. The X-ray contours 
illustrate levels  that are 10 per cent and 1 per cent of the maximum value in the map. The 
gross features are similar in all 3 models for both clusters, although the X-ray maps reveal
that the gas in the the PC clusters is the smoothest, while the GO clusters contain
gas with the most small-scale structure.}
\label{fig:clusmapz0}
\end{figure*}

We also compute the thermal SZ effect due to an individual cluster by
constructing Compton-$y$ maps. This allows us to separate the cluster contribution
(within a cylinder) from the total integrated signal along the line-of-sight.  
Each map is constructed by first identifying all hot gas particles within a 
cuboid of size $2r_{\rm vir} \times 2r_{\rm vir} \times 6r_{\rm vir}$, 
centred on the cluster. The particles are then projected along
the long axis of the cuboid and smoothed on to a 2D grid, creating
the $y$ distribution. We estimate $y$ at the location of each pixel, 
$\mathbf{R}_{\rm p}=(x,y)$, as
\begin{equation}
y(\mathbf{R}_{\rm p}) = 
{\sigma_{\rm T} k m_{\rm gas} \over A_{\rm pix} \mu_{\rm e} m_{\rm H} m_{\rm e} c^2 }
\sum_i \, {w(|\mathbf{R}_{i}-\mathbf{R}_{\rm p}|,h_i) T_i \over
  \sum_{\rm p} \, w(|\mathbf{R}_{i}-\mathbf{R}_{\rm p}|,h_i) },
\label{eqn:ymap}
\end{equation}
where $A_{\rm pix}$ is the area of a single pixel and $w$ is the
projected version of the SPH kernel used by {\sc gadget2}. 
The main sum runs over all hot gas particles with projected position $\mathbf{R}_{i}$, 
temperature $T_i$ and SPH smoothing length $h_i$.
The sum in the denominator runs over all pixels and normalises the
kernel for each particle.

Fig.~\ref{fig:clusmapz0} illustrates Compton-$y$ maps for
two massive clusters in our simulations at $z=0$: a regular ($s \simeq 0.02$) 
cluster with a virial mass $M_{\rm vir} \simeq 2.9\times 10^{15}\hMsol$ (the most 
massive object in the MS) and a merging ($s \simeq 0.1$) 
cluster with $M_{\rm vir} \simeq 1.5\times 10^{15}\hMsol$. The left panels show 
results for the GO simulation, the middle panels for the PC simulation and the 
right panels for the FO simulation.  

As has been seen in previous simulations (e.g. \citealt{Motl05}), 
the $y$ distribution is very smooth. 
The most significant features are sharp edges associated with shocks; this is 
especially clear in the case of the merging cluster. Qualitatively, the maps look 
structurally similar between models although their $y$ values within a given pixel 
can be significantly different, with the GO and PC models lying at either
extremes. For the regular GO cluster, the mean $y$ within the
virial radius is $\left< y \right> = 1.4 \times 10^{-5}$, with a range of values from 
$2\times 10^{-7}-7\times 10^{-4}$. For the PC cluster, the mean value is very similar 
although the maximum $y$ (associated with the centre of the cluster) is almost half 
($y_{\rm max} = 4 \times 10^{-4}$). This is due to the pre-heating of the gas   
that acts to smooth out the high density regions. 

It is also noticeable that the
GO clusters contain a significant amount of small-scale structure in the gas. This is
not clear in the $y$ distribution but is evident from the overlaid X-ray surface 
brightness contours.
\footnote{X-ray surface brightness maps are calculated by replacing $T_i$ in 
equation~(\ref{eqn:ymap}) with $\rho_i \Lambda(T_i,Z)$, where $\rho_i$ is the density
of hot gas particle $i$ and $Z=0.3Z_{\odot}$ is the assumed metallicity. The 
cooling function, $\Lambda(T,Z)$, is calculated for the soft [0.5-2] keV band. We
normalise each surface brightness map to the maximum pixel value.}
These are clumps of low entropy gas associated with 
substructures in the cluster. Again, the pre-heating has smoothed these out by raising
the entropy of the gas at high redshift. These features are also seen in the FO clusters,
where heating is localised to haloes in which AGN feedback is occurring.

\subsection{Sky maps}
\label{sec:skymaps}

\begin{figure*}
\centering
\includegraphics[width=8.5cm]{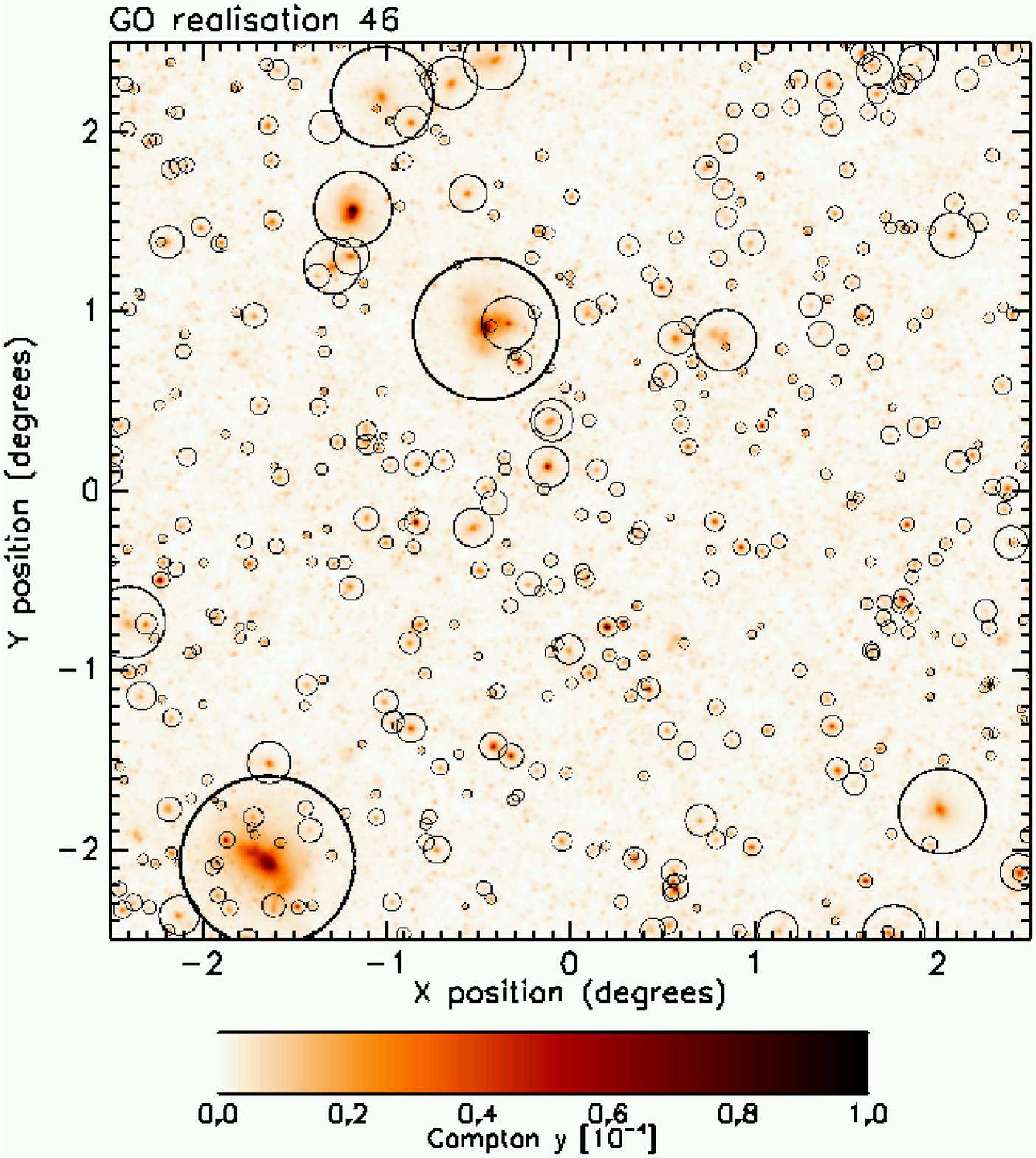}
\includegraphics[width=8.5cm]{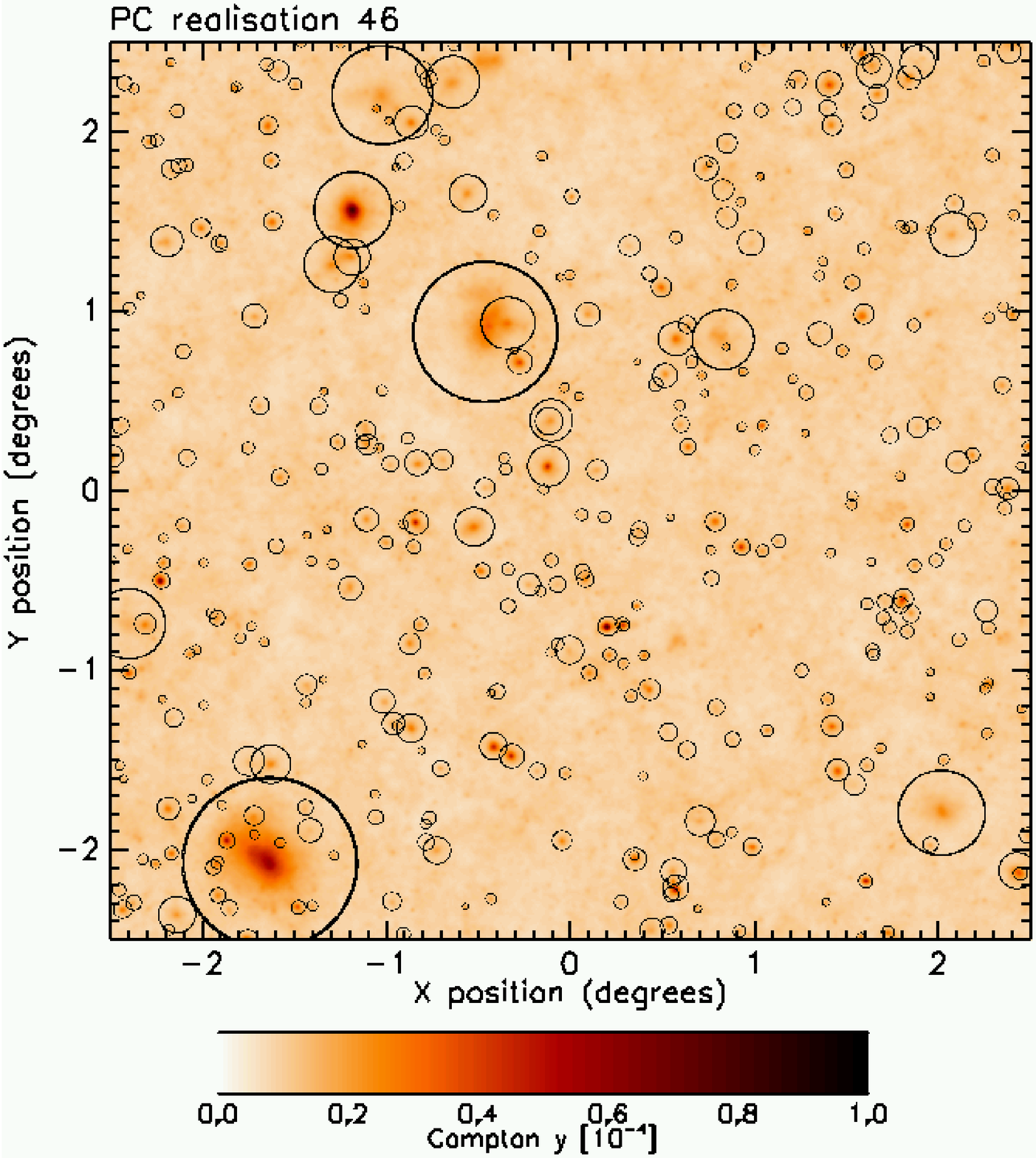}
\caption{An example $5\times 5$ square degree sky map of the 
Compton $y$ parameter from the GO (left) and PC (right) simulations.
The maps were smoothed with a Gaussian kernel with a full-width half-maximum
of 1 arcmin. Circles
illustrating virial radii of all clusters with $M_{\rm vir}>10^{14}\hMsol$ are overlaid. 
Although both maps
contain the same clusters, the PC map clearly has a larger background and the $y$ distribution
within each cluster is smoother, as seen in the maps of individual objects.}
\label{fig:ymaps}
\end{figure*}

\begin{figure}
\centering
\includegraphics[width=8cm]{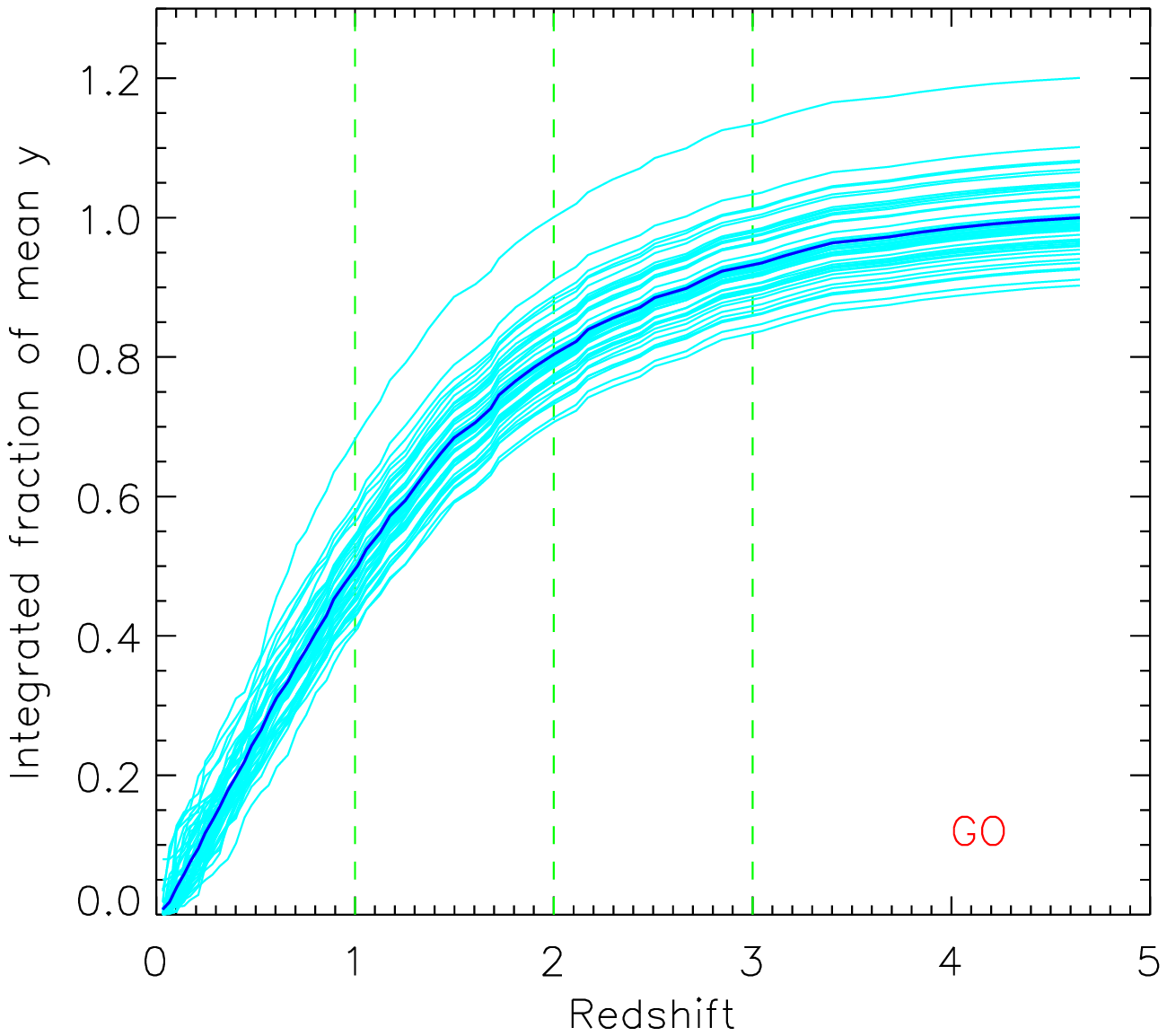}
\includegraphics[width=8cm]{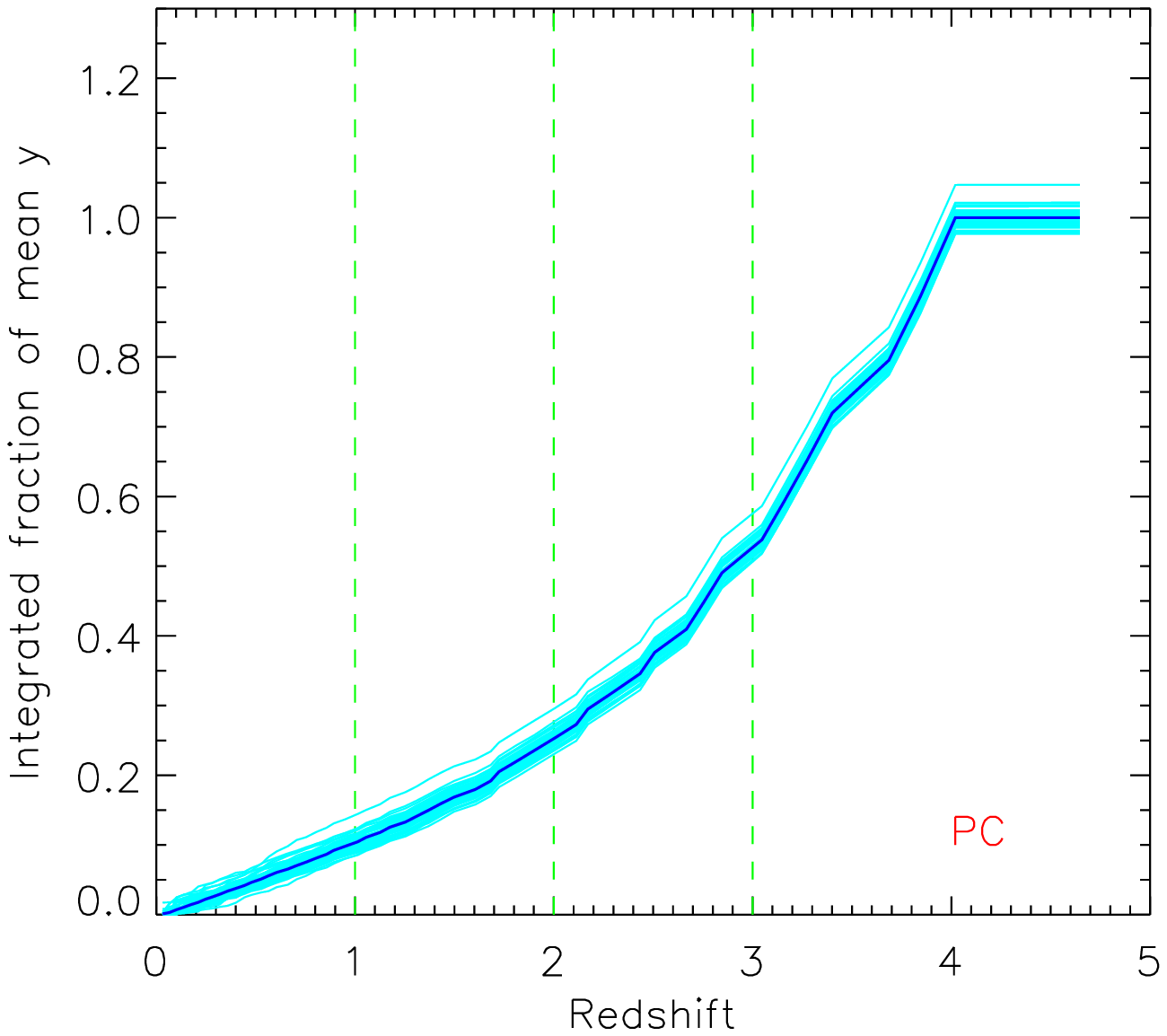}
\caption{Integrated contribution to the overall mean $y$ signal from gas below
a given redshift. The top panel is for the GO simulation and the bottom panel for the PC simulation.
Light curves are for individual maps and the dark curve is the average over all 50 maps. Each curve
is normalised to the mean $y$ averaged over all 50 maps, highlighting the scatter in the integrated
signal between realisations. The simulations
predict dramatically different redshift dependencies: the mean signal in the GO simulation comes from
low redshfit ($z<2$) whereas the opposite is the case for the PC simulation, due to the effect of 
pre-heating at $z=4$.}
\label{fig:cumumeany}
\end{figure}

We also analyse simulated sky maps of the thermal SZ effect for the
GO and PC models, using the {\it stacked box} approach pioneered by 
\citet{daSilva00}. This is an approximate method for
generating past light-cones using a finite number of outputs. To do this
we first compute the lookback time corresponding to a comoving distance
of $50 \hMpc$. We then calculate successive lookback times, increasing
the comoving distance in steps of $\Delta_{\rm map}=100 \hMpc$. These
lookback times are used to find the nearest output time when
simulation data are stored (a total of 160 snapshots were generated). 
We also calculate the comoving width required at each lookback time, 
corresponding to a fixed opening angle of $\theta_{\rm map}=5^{\circ}$.
The final lookback time is chosen such that the comoving width is still
smaller than the box-size, to avoid replication of the particles. The
choice of $\theta_{\rm map}$ allows us to integrate the SZ effect out
to a maximum redshift, $z_{\rm max}=4.7$, using 47 snapshots; this is 
sufficiently large for the mean $y$ signal to be converged in our 
simulations (see Fig.~\ref{fig:cumumeany}, discussed below). 

Once the required volumes are defined to make up the lightcone, the 
second stage is to use a random number generator to construct a table of random translations, 
rotations (in steps of $\pi/2$ radians) and reflections about each of the
three axes. This is done in order to minimise the chance of the lightcone
containing the same cluster at different redshifts (note the volume required
at each time is always less than 20 per cent of the full simulation box because
of our choice of $\Delta_{\rm map}$). The list of operations are then used
to determine which particles are required to compute the contribution to the
SZ signal from each redshift (used to create a so-called {\it partial} map)
This stage is repeated 50 times to allow us to generate 50 quasi-independent realisations. 

The final stage is to generate the partial maps themselves, by smoothing the 
appropriate gas particles on to a 2D grid. This is done using
the same technique as for individual clusters but now using a map area corresponding to 
$\theta_{\rm map} \times \theta_{\rm map}$ at each redshift and a comoving depth
of $\Delta_{\rm map}$. Each partial map contains $1200 \times 1200$ pixels such that
each pixel has an angular size, $\theta_{\rm pix}=0.25$ arcmin, comfortably smaller
than the typical resolution of current SZ telescopes (1-10 arcmin).
The 47 partial maps are then stacked for each realisation to make final maps of the $y$ parameter.

Fig.~\ref{fig:ymaps} shows an example Compton-$y$ sky map for realisation 46, chosen
because it contains a relatively large cluster. Both the GO (left panel) and PC (right panel)
versions are shown. The maps were smoothed using a Gaussian kernel with 
a full-width half-maximum of 1 arcmin, similar to the resolution of 
modern ground-based SZ telescopes such as SPT and ACT.

The most striking difference between the two maps is the contrast: the PC map has a higher 
background than the GO map, making it harder to visually pick out the SZ sources 
associated with the clusters. This is due to the extra thermal energy added to {\it all} 
the gas by the pre-heating process and can be quantified by measuring the mean $y$ parameter,
averaged over all 50 realisations. For the GO run, we find $\left< y \right> = 2.3\times 10^{-6}$, increasing
by more than a factor of four to $\left< y \right> = 9.9 \times 10^{-6}$ for the PC run. Although
both values are below the current constraint from {\it COBE}/FIRAS, 
$\left< y \right> < 1.5\times 10^{-5}$
\citep{Fixsen96}, it is unlikely to be the case that the true background is
as high as in the PC model, as this would erase many of the weak neutral 
hydrogen absorption lines seen towards quasars \citep{Theuns01,Shang07,Borgani09}. The PC model
therefore serves as an extreme test of the effect of a high background although we
will remove the mean $y$ signal in our analysis in Section~\ref{subsec:projection} 
to mimic observations.

The contribution to the mean $y$ signal from gas at different redshifts is shown in
Fig.~\ref{fig:cumumeany}. The top panel shows results for all 50 maps
in the GO simulation and the bottom panel for the PC simulation. Again, the difference 
between the two models is striking: the majority of the $y$ signal comes from 
low redshift in the GO model (around 80 per cent from $z<2$) whereas
the opposite is true for the PC model (around 80 per cent from $z<3.5$). Most of the
mean $y$ comes from overdense regions (groups and clusters) in the GO model that
are more abundant at low redshift. In the PC case, most of the mean signal comes from
mildly overdense gas at high redshift \citep{daSilva01}. Note also that the contribution from
gas at $z>4$ is approximately zero in the PC model, unlike in the GO case, where there is a
small but non-negligible signal. This difference is due to the inclusion of radiative cooling
in the former model which removes most of the (small amount of) ionised gas at these redshifts.
%\vspace{-0.5cm}

\section{Hot gas pressure profiles}
\label{sec:pressureprofiles}

Fundamental to understanding the SZ effect from clusters is the 
hot gas pressure profile, since we can write the SZ $Y$ parameter for
a spherically symmetric cluster as
\begin{equation}
Y_{500} = 
{\sigma_{\rm T} \over  m_{\rm e} c^2} \, \int_{0}^{r_{500}} \, P_{\rm e}(r) \, 4\pi r^3 {\rm d}\ln r,
\end{equation}
where $P_{\rm e}=n_{\rm e}kT_{\rm e}$ is the electron pressure. The contribution to $Y_{500}$
will therefore be highest at the radius where $r^3 P_{500}$ is maximal. If the gas is in hydrostatic 
equilibrium then the pressure profile ought to be structurally similar between different
clusters since it is directly constrained by the underlying gravitational potential, 
which itself takes on a regular form (e.g. \citealt{Navarro97}, hereafter NFW). 

We construct and compare spherically-averaged,
hot gas mass-weighted pressure profiles using equation~(\ref{eqn:pprof}),
for all clusters with 
$M_{500}>10^{14}\hMsol$ in our three (GO, PC and FO) models at $z=1$ and $z=0$. The profiles
are re-scaled such that we plot dimensionless quantities $x^3 P(x)/P_{500}$ against 
$x$, where $x=r/r_{500}$ and the scale pressure, $P_{500} \propto M_{500}^{2/3} E(z)^{8/3}$, is 
determined assuming a self-similar isothermal gas distribution \citep{Voit05}. If clusters
formed a self-similar population then these re-scaled profiles would be identical for 
both varying mass and redshift.

\begin{figure}
\centering
\includegraphics[width=8.5cm]{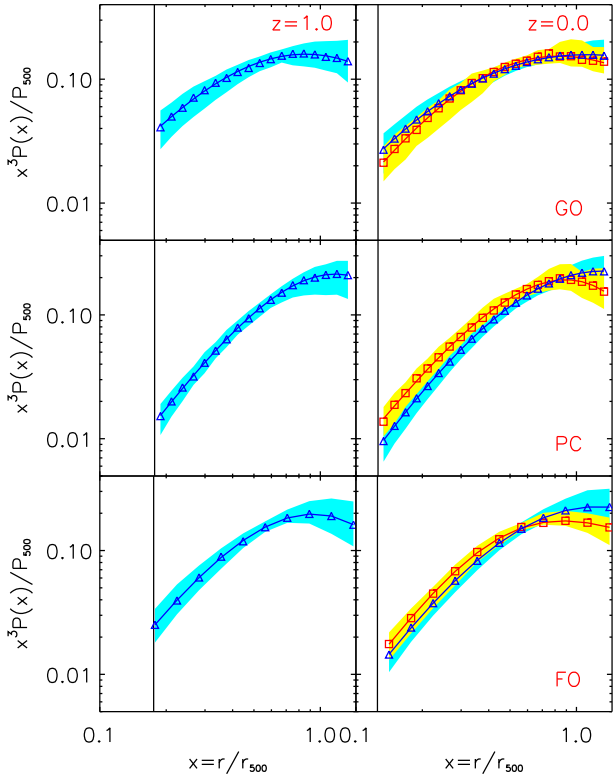}
\caption{Scaled pressure profiles for clusters within the 
GO (top panels), PC (middle panels) and FO (bottom panels) 
simulations. Left panels are $z=1$ results and right panels for $z=0$.
Median profiles are shown separately for low-mass 
($10^{14}\hMsol < M_{500} \leq 5\times 10^{14}\hMsol$; triangles)
and high-mass ($M_{500}>5\times 10^{14}\hMsol$; squares) respectively
(note there are no high mass clusters at $z=1$).
The yellow (cyan) shaded band illustrates the 16/84 per centiles (and thus represents the 
cluster-to-cluster scatter) for the high-mass (low-mass) sub-sample. 
Solid curves are best-fit generalised NFW profiles to the median pressure profiles.
The vertical solid line represents the radius where gravity is softened in
the cluster with the smallest $r_{500}$ (this is at a smaller radius than plotted 
for the FO model, for which a smaller softening was used,
but we choose to use the same scale as in the other two models for ease of comparison). 
Note that the contribution to $Y_{500}$ is predominantly from $r>0.5 r_{500}$ and so
is not particularly susceptible to variations in the cluster core.
}
\label{fig:preprofz}
\end{figure}

Median scaled profiles are shown in Fig.~\ref{fig:preprofz}, split into low-mass 
($10^{14}\hMsol < M_{500} \leq 5\times 10^{14}\hMsol$; triangles) and high-mass
($M_{500}>5\times 10^{14}\hMsol$; squares) sub-samples. Comparing the high-mass
clusters between the three models at $z=0$, it is immediately apparent that the 
largest contribution to $Y_{500}$ comes from radii close to $r_{500}$, i.e. where
$P(r) \propto r^{-3}$. The profiles rise sharply (by around an order of magnitude) 
from the core outwards then stay level or gradually decline at larger scales. 
The largest differences between the three models occur in the core region, where 
the PC and FO clusters have lower central pressures than the GO clusters due to 
the increase in core gas entropy from the extra heating.

The low-mass clusters have very similar median profiles to the high-mass clusters 
in the GO simulation, reflecting the similarity of objects in that model \citep{Stanek10}. 
In the PC and FO models however, the pressure profiles of the low-mass clusters 
have markedly different shapes from their high-mass counterparts. In particular, the
scaled pressure in low-mass clusters is lower in the central region and is higher in the outer 
region, indicating that they are less concentrated than the high-mass clusters. 
Again this reflects the breaking of self-similarity caused by the feedback/pre-heating 
which has a larger effect in the lower mass clusters; the extra entropy given to the gas 
causes a re-distribution to take place, pushing the gas out to larger radius. 

Comparing the low-mass clusters at low and high redshift, the GO model shows 
little evolution (the core pressures are slightly lower), while clusters 
in the PC model have significantly lower core pressures at $z=1$. This reflects the larger 
impact of the pre-heating on the gas at high redshift, since a cluster of fixed mass has a 
lower characteristic entropy at higher redshift from gravitational heating 
[$K \sim M^{2/3} E(z)^{-2/3}$]. Interestingly, the scaled pressure
profiles in the FO model show little evolution with redshift, although the pressure in the 
outskirts $(r>r_{500})$ is higher at $z=0$, reflecting the late-time heating of the gas by AGN.

\begin{figure}
\centering
\includegraphics[width=8.5cm]{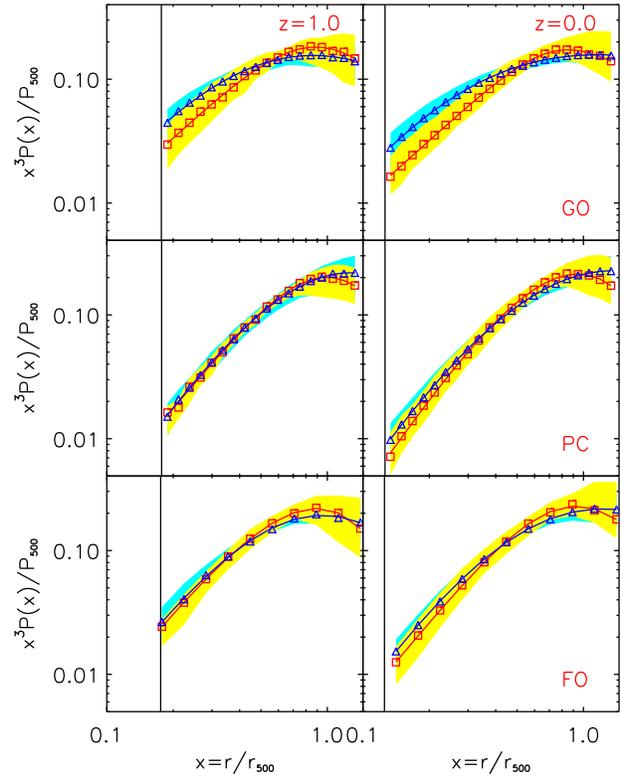}
\caption{As in Fig.~\ref{fig:preprofz} but the samples are now split into 
regular ($s\le 0.1$; triangles) and disturbed ($s> 0.1$; squares) sub-samples. Disturbed
(merging) clusters tend to have lower central pressures but higher peak values of $x^3 P(x)$.}
\label{fig:preprofzsub}
\end{figure}

We also compare scaled pressure profiles between regular ($s\le 0.1$) and disturbed ($s>0.1$)
clusters in Fig.~\ref{fig:preprofzsub}, for our samples with $M_{500}>10^{14}\hMsol$. The largest
differences between the two sub-samples can be seen for the GO model, where the disturbed clusters
(squares) have lower scaled pressure everywhere except around the maximum at $r \simeq 0.9 r_{500}$. 
This is because the ongoing merger is compressing the gas (and therefore increasing its pressure) at 
large radius while the inner region has yet to respond to the increase in the mass of the system. 
Note that since $Y_{500}$ is proportional to the area under the pressure profile, there will be a noticeable
offset in the $Y_{500}-M_{500}$ relation, where a disturbed cluster has a smaller $Y_{500}$ than a
regular cluster with the same mass (see the next section). 
These differences are still present but at a lower level in the
PC and FO models, where the higher entropy of the gas in lower-mass clusters means that it is less easily
compressed. This in turn leads to a negligible 
offset between regular and disturbed clusters in the $Y_{500}-M_{500}$
relation, as we will show in the next section.

\begin{figure}
\centering
\includegraphics[width=8.5cm]{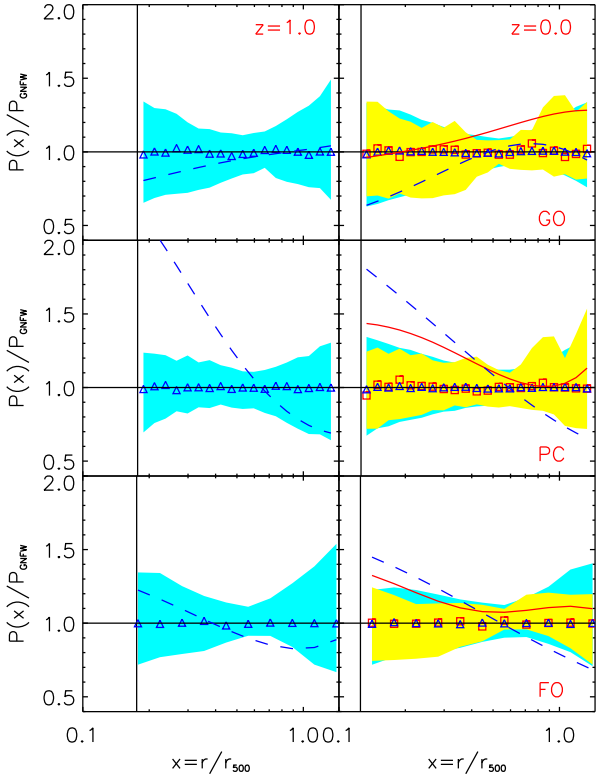}
\caption{As in Fig.~\ref{fig:preprofz} but the median pressure profiles (and
scatter) are now shown relative to their best-fit generalised NFW model, 
clearly showing the size of cluster-to-cluster variations (that can be as large 
as 50 per cent). 
Solid and dashed curves are observed mean pressure profiles from low-redshift 
X-ray data (REXCESS; \citealt{Arnaud10}), again scaled to our best-fit 
generalised NFW model profiles, assuming the median mass from our low and
high-mass sub-samples respectively. In the outer regions ($r>0.5 r_{500}$) 
the high-mass clusters in the PC and FO models fit the latter profile quite 
well (to within 10 per cent or so), but the difference is larger for 
low-mass clusters, especially in the core regions.}
\label{fig:preprofrelz}
\end{figure}

The shaded bands in Figs.~\ref{fig:preprofz} and \ref{fig:preprofzsub} illustrate the 16/84 per centiles for 
the two respctive sub-samples and thus gives an indication of the cluster-to-cluster scatter. We show this
more clearly in Fig.~\ref{fig:preprofrelz}, where we have normalised the clusters in the 
low and high mass sub-samples to the generalised NFW model that best fits the median profile (see below). 
Although the scatter at fixed radius is quite low compared with some other properties such as X-ray 
surface brightness, it is nevertheless appreciable and can be as high as 30-50 per cent beyond $r_{500}$.
Thus it is clearly not accurate to assume a single profile to describe all clusters, especially around
$r_{500}$ and beyond, where much of the SZ signal comes from.

\subsection{Generalised NFW model}

In a previous study of hot gas pressure profiles in cosmological simulations,
\citet{Nagai07b} found that the mean pressure profile of their simulated
clusters could be well described by a generalised NFW (GNFW) model with five free parameters
\begin{equation}
{P(r) \over P_{500}} = {P_0 \over u^{\gamma} (1+u^{\alpha})^{(\beta-\gamma) / \alpha}},
\label{eqn:gnfw}
\end{equation}
where $u=c_{500}x$, $c_{500}$ is the concentration parameter, $P_0$ the normalisation parameter
and $(\gamma, \alpha,\beta)$ determine the shape of the profile at small 
($u \ll 1$), intermediate ($u \simeq 1$) and large ($u \gg 1$) radius respectively. The GNFW
model has been shown to provide a good description to the pressure profiles of X-ray groups and clusters 
(e.g. \citealt{Arnaud10,Sun11}) and is being used to optimise SZ cluster detection in data from the
{\it Planck} satellite (e.g. \citealt{Planck11c}).

\begin{table}
\caption{Best-fit parameters for the generalised NFW model when applied to
our median hot gas pressure profiles. Column 1 gives the redshift; column 2
the simulation model and cluster sub-sample (LM and HM refer to the low and
high-mass sub-samples, respectively); and columns 3-7 the parameter values 
(see text for their meanings).}
\begin{center}
\begin{tabular}{llccccc}
\hline
Redshift & Clusters & $P_{0}$ & $c_{500}$ & $\gamma$ & $\alpha$ & $\beta$\\
\hline
$z=0$ & GO/LM & 33.788 & 2.925 & 0.267 & 0.944 & \phantom{-}1.970\\
      & PC/LM & \phantom{-}6.317 & 0.517 & 0.090 & 0.901 & \phantom{-}1.603\\
      & FO/LM & \phantom{-}4.732 & 1.052 & 0.298 & 1.108 & \phantom{-}2.371\\
\cline{2-7}
      & GO/HM & \phantom{-}6.756 & 1.816 & 0.519 & 1.300 & \phantom{-}2.870\\
      & PC/HM & \phantom{-}0.938 & 0.183 & 0.584 & 1.114 & 11.885\\
      & FO/HM & \phantom{-}3.210 & 1.974 & 0.605 & 2.041 & \phantom{-}2.989\\
\cline{2-7}
$z=1$ & GO/LM & 11.994 & 0.700 & 0.345 & 0.837 & \phantom{-}3.610\\
      & PC/LM & \phantom{-}0.856 & 0.539 & 0.512 & 1.447 & \phantom{-}4.038\\
      & FO/LM & \phantom{-}2.734 & 0.349 & 0.375 & 1.055 & \phantom{-}5.049\\
\hline
\end{tabular}
\end{center}
\label{tab:gnfw}
\end{table}

We have applied the GNFW model to our simulated clusters and the results for the median profiles 
can be seen as solid curves in Figs.~\ref{fig:preprofz} and \ref{fig:preprofzsub}. We also normalise
our pressure profiles to the best-fitting median GNFW profile in Fig.~\ref{fig:preprofrelz}. The 
residual values for our median profiles are also shown (as triangles and squares for our low and high 
mass sub-samples) and are clearly at the per cent level. Such small residuals are not surprising 
given the model contains five free parameters (once $r_{500}$ is specified). The best-fit parameter
values themselves are listed in Table~\ref{tab:gnfw}.

\begin{figure*}
\centering
\includegraphics[width=16cm]{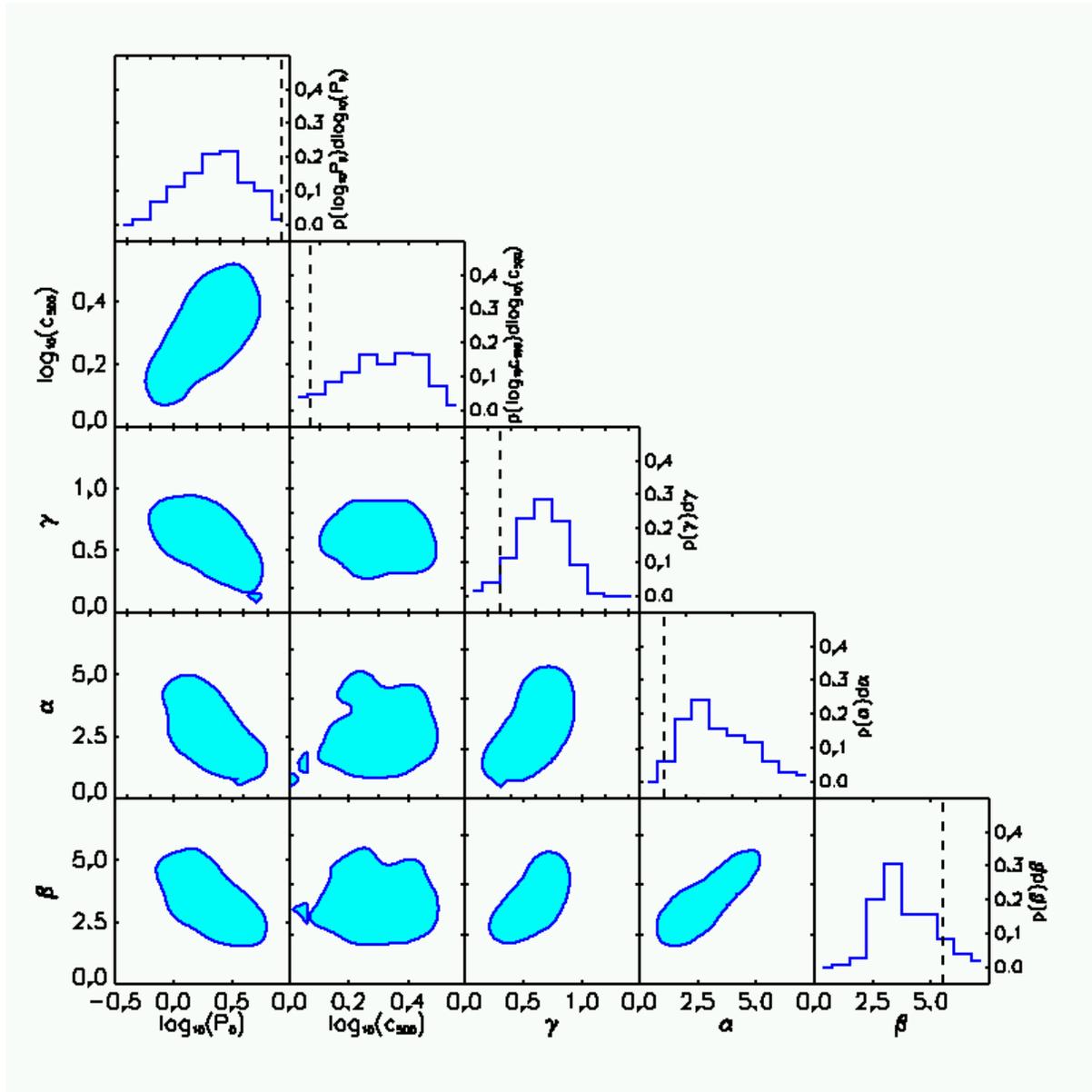}
\caption{Estimated likelihood distributions for the best-fit generalised NFW 
model parameters that describe the hot gas pressure profiles in the FO
simulation at $z=0$. The off-diagonal panels show 68 per cent confidence 
regions for the marginalised 2D distributions, for all parameter combinations.
The diagonal panels show the 1D marginalised distributions for each parameter,
with the best-fit parameter values from REXCESS data overlaid as 
dashed vertical lines. The three slope parameters ($\alpha,\beta,\gamma$) are
strongly correlated with one another and are all negatively correlated with the
normalisation, $\log_{10} P_0$. The concentration parameter, 
$\log_{10} c_{500}$, is only correlated with the normalisation.}
\label{fig:gnfwparams}
\end{figure*}

To investigate the distribution of GNFW parameters and any degeneracies that arise between
parameters, we plot marginalised likelihood distributions for the FO model at $z=0$ 
in Fig.~\ref{fig:gnfwparams}. The full five-dimensional likelihood distribution is estimated
by fitting the GNFW model to individual clusters and computing the frequency of parameters
$[\log_{10} P_0, \log_{10} c_{500}, \gamma, \alpha, \beta]$ over a five-dimensional grid, 
which is then normalised such that the sum over all allowed parameter values is unity. We 
assume, as prior information, that the allowed range for each parameter is as specified on
the axes in Fig.~\ref{fig:gnfwparams} and that each value is equally likely.  

The diagonal panels in Fig.~\ref{fig:gnfwparams} show the marginalised 1D likelihood 
distributions for each of the five parameters, while the off-diagonal panels show the
68 per cent confidence regions for the full range of marginalised 2D distributions, 
smoothed to reduce noise. The concentration parameter is strongly correlated with the 
normalisation parameter but does not correlate strongly to any of the slope parameters. 
Interestingly, the normalisation is anti-correlated (and therefore degenerate) with the 
slope parameters. Finally, the slope parameters are strongly correlated with one another.
It is therefore clear that a simpler model with fewer slope parameters could potentially 
be found that describe these simulated data. However, the flexibility of the GNFW model 
allows a wide range of profiles to be accurately described using a simple formula.
This is especially true when cool-core clusters are included; these are absent in our
current models and so we plan to extend our work to investigate cooling effects in a 
future study. 

\subsection{Comparison with observational data}

We also compare the simulated profiles with the pressure profile
presented by \citet{Arnaud10}, compiled from low-redshift X-ray observations (for $r<r_{500}$; 
the REXCESS sample) and other numerical simulations (for $r>r_{500}$). It therefore provides
information on the realism of our simulated pressure profiles as well providing a useful 
comparison with other simulations (on large scales).

The \citet{Arnaud10} profile is based on the GNFW model, modified to account for additional (weak) mass dependence
in the observational data
\begin{equation}
P(x) = P_{\rm GNFW} \, \left[ {M_{500} \over 3\times10^{14} \, {\rm M}_{\odot} } \right]^{\alpha_{\rm P}},
\end{equation}
where $P_{\rm GNFW}$ is the GNFW pressure profile given in equation~(\ref{eqn:gnfw}) with parameters, 
$[P_0,c_{500},\gamma,\alpha,\beta]=[8.403,1.177,0.3081,1.0510,5.4905]$ and $\alpha_{\rm P}=0.12$.
We show this profile, evaluated for the median mass values of our two sub-samples, 
in Fig.~\ref{fig:preprofrelz}; the dashed curve is for the low-mass sample, plotted relative to 
our best-fit GNFW profile, while the solid curve is for the high-mass sample. The Arnaud et al.
parameters are also shown as dashed lines in Fig.~\ref{fig:gnfwparams}.

Comparing with our $z=0$ results, as is most appropriate, the median GO profiles
agree to within 30 per cent or so, over the plotted range of radii and for both mass ranges. 
For the PC and FO clusters, the agreement is very good at large radius ($>0.5 r_{500}$) for high-mass clusters, 
where the Arnaud et al. profile is only around 10 per cent higher and within the intrinsic
scatter of our simulated profiles.
The low-mass clusters are more discrepant, with the steeper Arnaud et al. profile being 
20-30 per cent lower at $r_{500}$. This suggests that our simulated clusters  contain
gas that is at higher pressure at $r_{500}$ than in those used for the Arnaud et al. profile
at large radius. Given that the feedback in our models is likely to be stronger than in the simulations
used in the Arnaud et al. study, this discrepancy in pressure is probably due to the effects of 
radiative cooling, absent in our models and likely significant in the other simulations (see the discussion
in Section~\ref{subsec:ymsim}; we also note that Arnaud et al. already corrected for the effects of baryon fraction). 
Even larger differences are present in the inner regions; there, the Arnaud et al. profile is significantly 
higher than our simulated results. Again, cooling is the likely culprit here as its effect is strongest 
in the densest regions.

An important uncertainty in the observed profile estimation is the effect of hydrostatic bias, i.e. 
systematic offsets in $r_{500}$ and $M_{500}$ from their true values, when estimated 
from the equation of hydrostatic equilibrium. As we will show in Section~\ref{sec:hse}, 
hydrostatic mass is biased low with respect to the true mass and is most significant
for the GO model (the estimated-to true mass ratio is around 0.7 for the GO model, compared
with around 0.9 for the PC/FO models). The effect of this bias is to increase the scaled pressure
at fixed scaled radius, as both the scale radius, $r_{500}$, and the scale pressure, 
$P_{500} \propto M_{500}^{2/3}$, decrease, on average. We discuss the effect of hydrostatic bias 
on the $Y_{500}-M_{500}$ relation in detail in Section~\ref{sec:hse} but note here that we have 
explicitly checked how this affects the pressure profiles for each model. To do this, we 
first re-defined our sub-samples using the estimated masses. We then compared the shift in pressure 
at the estimated value of $r_{500}$ between the median scale profile and the Arnaud et al. profile, 
for both low-mass and high-mass sub-samples. We also re-computed the pressure profiles using the
spectroscopic-like temperature, rather than the hot gas mass-weighted temperature, as this will be 
closer to the X-ray temperature profile used by Arnaud et al. 

We find that the combined effect of these changes is largest for the GO model, 
where the median pressure profiles from both sub-samples are now within 10 per cent of 
the Arnaud et al. values at $r_{500}$. The increase in the scaled pressure profile due to 
hydrostatic bias is counteracted by a decrease due to the use of spectroscopic-like temperature, 
which is lower than the mass-weighted temperature for this model (see Section~\ref{sec:yobs}).
The two effects are smaller for the PC and FO models and so we see very similar results to those 
before these changes were applied. Thus, the scaled pressure profiles for the low-mass clusters 
in these models are still around 30 per cent lower than the Arnaud et al. profile at $r_{500}$.
%\vspace{-0.5cm}

\section{SZ scaling relations}
\label{sec:scalingrelations}

\begin{table*}
\caption{Best-fit parameters for simulated SZ scaling relations at $z=0$ and $z=1$. 
Column 1 gives the scaling relation being considered; 
column 2 the pivot point (in appropriate units); 
column 3 the redshift; 
column 4 the simulation model;
and columns 5-7 the best-fit values for the normalisation, power-law index and
scatter in $\log_{10}Y_{500}$ respectively. Quoted uncertainties correspond to either
the 16th or 84th per centile (whichever is largest), estimated using the bootstrap 
re-sampling technique.}
\begin{center}
\begin{tabular}{llllccc}
\hline
Relation & $X_0$ & Redshift & Model & $A$ & $B$ & $\sigma_{\log_{10}Y}$\\
\hline
$E(z)^{-2/3} Y_{500}-M_{\rm 500}$ & $3\times 10^{14}\hMsol$ & $z=0$ &
    GO & $-4.754 \pm 0.002$ & $1.670 \pm 0.007$ & $0.041 \pm 0.001$ \\
&&& PC & $-4.774 \pm 0.003$ & $1.794 \pm 0.009$ & $0.045 \pm 0.001$ \\
&&& FO & $-4.744 \pm 0.003$ & $1.69  \pm 0.02$  & $0.043 \pm 0.003$ \\
\cline{4-7}
&& $z=1$ & 
     GO & $-4.79 \pm 0.01$ & $1.60 \pm 0.04$ & $0.048 \pm 0.003$ \\
&&& PC & $-4.82 \pm 0.01$ & $1.83 \pm 0.04$ & $0.059 \pm 0.005$ \\
&&& FO & $-4.75 \pm 0.01$ & $1.63 \pm 0.04$ & $0.037 \pm 0.004$ \\
\hline
$E(z)^{-2/3} Y_{500}-M_{\rm gas,500}$ & $3\times 10^{13}\hMsol$ & $z=0$ &
    GO & $-5.098 \pm 0.001$ & $1.650 \pm 0.007$ & $0.029 \pm 0.001$ \\
&&& PC & $-4.887 \pm 0.001$ & $1.478 \pm 0.008$ & $0.018 \pm 0.001$ \\
&&& FO & $-4.889 \pm 0.003$ & $1.45  \pm 0.01$  & $0.025 \pm 0.002$ \\
\cline{4-7}
&& $z=1$ &
    GO & $-5.145 \pm 0.004$ & $1.61 \pm 0.03$ & $0.034 \pm 0.003$ \\
&&& PC & $-4.844 \pm 0.006$ & $1.46 \pm 0.05$ & $0.016 \pm 0.005$ \\
&&& FO & $-5.007 \pm 0.004$ & $1.53 \pm 0.03$ & $0.028 \pm 0.006$ \\
\hline
$E(z) Y_{500}-T_{\rm sl}$  & $5 \, {\rm keV}$ & $z=0$ &
    GO & $-4.27  \pm 0.03$  & $2.5  \pm 0.2$  & $0.19  \pm 0.02$  \\
&&& PC & $-4.706 \pm 0.006$ & $3.16 \pm 0.04$ & $0.060 \pm 0.002$ \\
&&& FO & $-4.665 \pm 0.006$ & $3.11 \pm 0.07$ & $0.078 \pm 0.007$ \\
\cline{4-7}
&& $z=1$ &
    GO & $-4.03 \pm 0.07$   & $3.0  \pm 0.6$  & $0.16  \pm 0.07$  \\
&&& PC & $-4.910 \pm 0.006$ & $3.38 \pm 0.06$ & $0.047 \pm 0.004$ \\
&&& FO & $-4.54 \pm 0.02$   & $2.8  \pm 0.1$  & $0.086 \pm 0.009$ \\
\hline
$E(z)^{-2/3} Y_{500}- E(z)^{-2/3} Y_{{\rm X},500}$ & $1\times 10^{-5}\YSZunits$ & $z=0$ &
    GO & $-4.952 \pm 0.003$ & $1.049 \pm 0.008$ & $0.058 \pm 0.002$ \\
&&& PC & $-5.020 \pm 0.001$ & $1.002 \pm 0.002$ & $0.015 \pm 0.001$ \\
&&& FO & $-5.012 \pm 0.001$ & $0.998 \pm 0.004$ & $0.018 \pm 0.001$ \\
\cline{4-7}
&& $z=1$ &
    GO & $-4.882 \pm 0.009$ & $1.05 \pm 0.03$ & $0.052 \pm 0.004$ \\
&&& PC & $-5.015 \pm 0.002$ & $0.999\pm 0.004$ & $0.009 \pm 0.001$ \\
&&& FO & $-5.007 \pm 0.003$ & $0.99\pm 0.01$ & $0.020 \pm 0.004$ \\
\hline
\end{tabular}
\end{center}
\label{tab:ymrelz0}
\end{table*}

We now present SZ scaling relations for our simulations and compare them specifically 
with the recent analysis of data from {\it Planck} and {\it XMM-Newton}. We will also compare our
results with recent simulations before going on to consider the effect of projection of 
large-scale structure along the line-of-sight. The effect of hydrostatic bias on the
scaling relations will be considered in the next section.

%The assumption of 
%hydrostatic equilibrium and its effect on SZ scaling relations is then investigated, 
%\subsection{True SZ Scaling Relations}

We consider the scaling relations between $Y_{500}$ and several 
other properties: the total mass, $M_{500}$; the hot gas mass, $M_{\rm gas,500}$;
the X-ray spectroscopic-like temperature, $T_{\rm sl}$; and the analagous X-ray 
quantity to $Y_{500}$, $Y_{\rm X,500}$. 
We note that the $Y_{\rm X}-M_{500}$ relation (not
considered here) has already been presented by \citet{Short10} and scaling relations 
for the lower density contrast, $\Delta=200$, for the GO and PC models by \citet{Stanek09}.
\footnote{We have independently verified that our GO and PC results, when using $\Delta=200$, 
are consistent with theirs,
but as was pointed out by \citet{Viana11} the $Y_{200}$ values given in \citet{Stanek09}
quoted with incorrect units.}

We follow the standard procedure and assume that the mean relationship 
between $Y_{500}$ and the independent variable 
can be adequately described by a power law and is thus a linear relationship
in log-space.
\footnote{\citet{Stanek10} present quadratic fits to the PC data but we find
this only to be important when the lower-mass groups are included, as was the
case in that study.}
We estimate
the slope and normalisation of the relation by performing a least-squares fit to the data
\begin{equation}
E(z)^{\gamma}Y_{500} = 10^{A} \, \left( X/X_{0} \right)^{B},
\label{eqn:ymrelz0}
\end{equation}
where $A$ and $B$ describe the best-fit normalisation and slope respectively and
$X_0$ is the pivot point, suitably chosen to minimise co-variance between the
two parameters. For the power-law index $\gamma$ we choose the appropriate value
for self-similar evolution, so if our clusters evolve self-similarly we should see
no change in the best-fit parameters $A$ and $B$.

We also estimate the scatter in $\log_{10}(Y_{500})$, $\sigma_{\log_{10}Y}$, as
\begin{equation}
\sigma_{\log_{10}Y} = \sqrt { {1 \over N-2} \, \sum_{i=1}^{N} \, 
\left[ \log_{10}Y_i  - \log_{10}Y_{\rm bf} (X_i) \right]^2 }, 
\end{equation}
where the index $i$ runs over all $N$ clusters included in the fit and 
$Y_{\rm bf}$ is the best-fit $Y_{500}$ value for a cluster with property, $X_i$. 
Note that the scatter in $\ln Y$ is simply $\sigma_{\ln Y} = \ln(10) \, \sigma_{\log_{10}Y}$.

\subsection{The $Y_{500}-M_{500}$ relation}

\begin{figure}
\centering
\includegraphics[width=8.5cm]{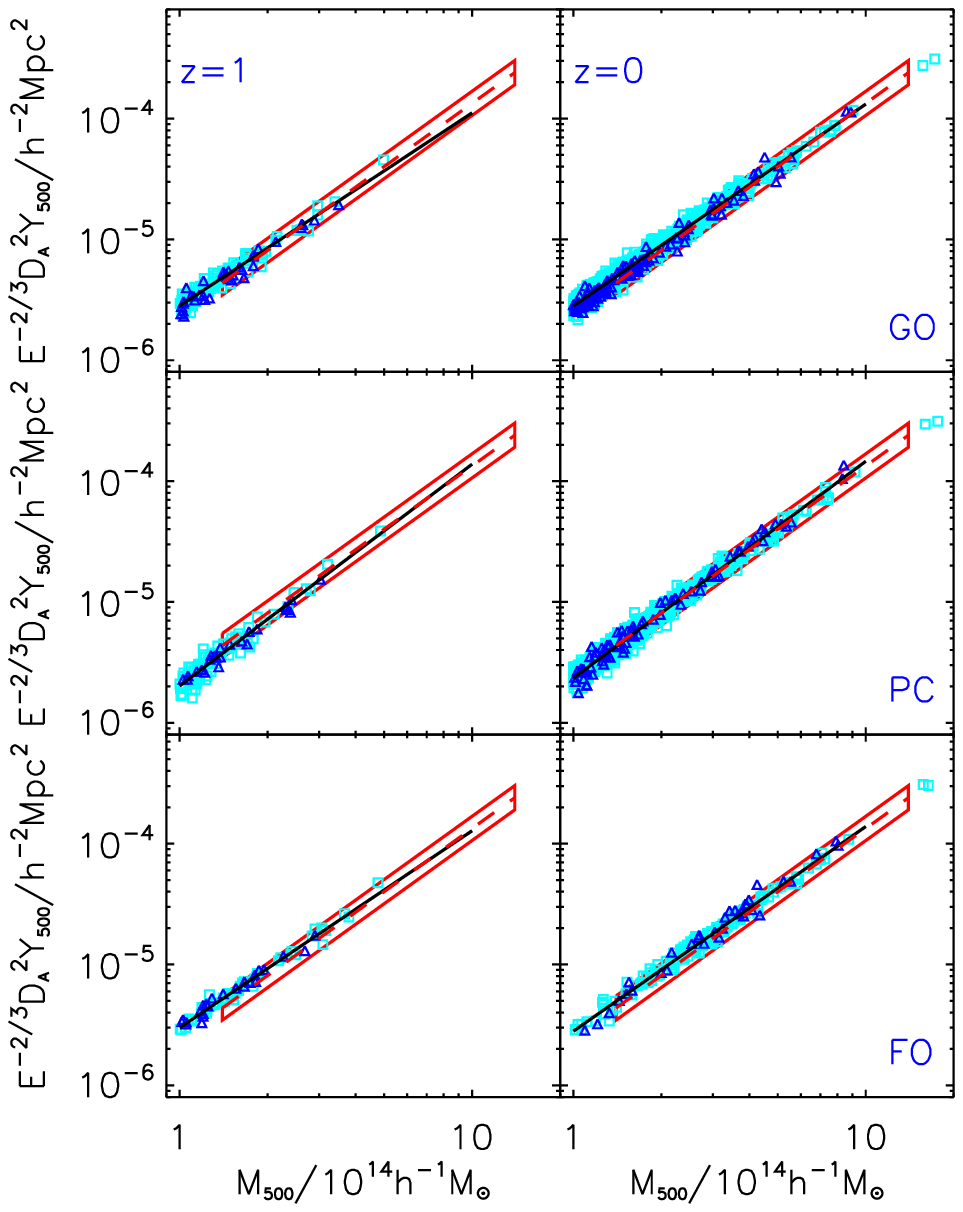}
\caption{
Scaling relations between the SZ flux, $Y_{500}$, and total mass, $M_{500}$,
for the GO (top panels), PC (middle) and FO (bottom) models at
$z=1$ (left) and $z=0$ (right panels). The $Y_{500}$ values at $z=1$ 
are re-scaled such that no change in the relation corresponds to self-similar
evolution ($Y_{500} \propto E(z)^{2/3}$ at fixed mass). The solid diagonal line
is a least-squares fit to the relation.  The best-fit power-law to  
$z<0.5$ {\it Planck}/{\it XMM-Newton} data \citep{Planck11c} is shown in all panels 
as a dashed line, while the box illustrates the intrinsic scatter in the observed
relation.
Triangles represent disturbed clusters (with $s>0.1$) while squares are regular
clusters. The results are very similar for all three models and there is 
no evidence for significant departure from self-similar evolution.}
\label{fig:ymrelz10}
\end{figure}

\begin{figure}
\centering
\includegraphics[width=8.5cm]{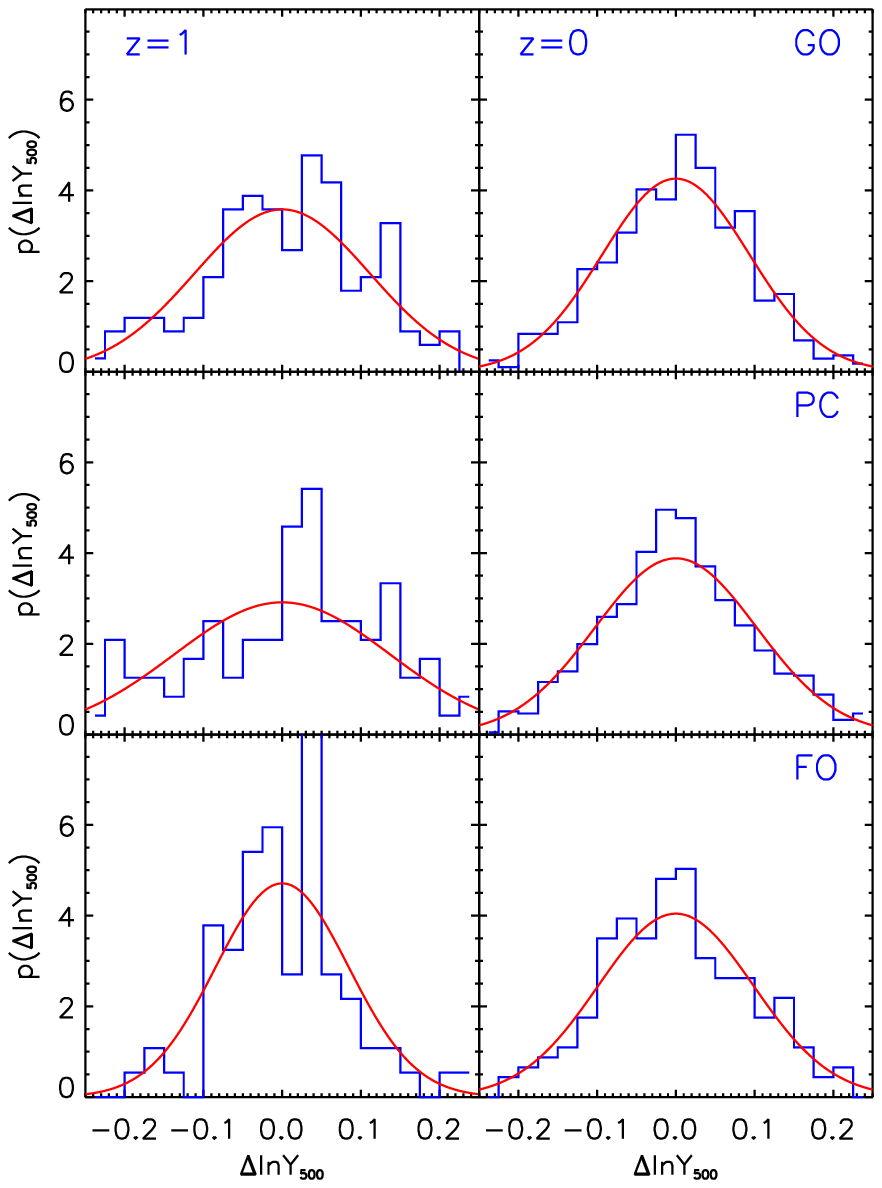}
\caption{
Distribution of residual $Y_{500}$ values about the best-fit $Y_{500}-M_{500}$ 
relation, plotted using natural logarithms, for each of the models at $z=1$ 
and $z=0$. A normal distribution of width $\sigma = \sigma_{\ln Y}$ is overlaid; 
it is clear that this provides a good description of the scatter.}
\label{fig:ymrelz10rdist}
\end{figure}

The most important scaling relation is that between SZ flux and mass. 
We present our $Y_{500}-M_{500}$ relations in Fig.~\ref{fig:ymrelz10}
for the GO model (top panels), PC model (middle panels) and FO model (bottom panels). 
Results are shown both for $z=1$ (left panels) and $z=0$ (right panels). 
The best-fit relation to all
clusters in each panel with $10^{14}\hMsol<M_{500}<10^{15}\hMsol$ is shown as a solid 
line (best-fit parameter values and their uncertainties are listed in Table~\ref{tab:ymrelz0}). 

It is clear that there is a very tight correlation between $Y_{500}$ and $M_{500}$
in all three models at both low and high redshift. At $z=0$ the intrinsic scatter
about the best-fit power law relation is only $\sim 4$ per cent, with sub-per cent
variations between models, making this particular relation one of the tightest known
cluster scaling relations involving gas; this finding is consistent with 
previous simulations with fewer clusters (e.g. \citealt{daSilva04,Nagai06}).
The distribution of residual $Y_{500}$ values about the best-fit relation is well 
described by a log-normal distribution of width $\sigma=\sigma_{\ln Y}$ (Fig.~\ref{fig:ymrelz10rdist}).
This is in agreement with previous work (e.g. \citealt{Stanek10,Fabjan11}).

The normalisation of the $z=0$ relation also varies
very little between models, the maximum variation being around 7 per cent. The best-fit slope also 
varies by around 7 per cent, from $1.67$ in the FO model (very close to the self-similar value of 
$5/3$) to $1.79$ in the PC model. As discussed in \citet{Short09} for the $Y_{\rm X}-M_{500}$ 
relation, the similarity between the models can be explained by the increase in gas temperature
compensating for the drop in gas mass, required to maintain virial equilibrium (since 
$Y \propto M_{\rm gas}T$ and is thus proportional to the total thermal energy of the gas).
The agreement between the GO and PC/FO models is better here than for the $Y_{\rm X}-M_{500}$
relation as $Y_{\rm X}$ is defined using the spectroscopic-like temperature, $T_{\rm sl}$, 
which is weighted more heavily by low entropy gas; we discuss this point further below.

\begin{figure*}
\centering
\includegraphics[width=18cm]{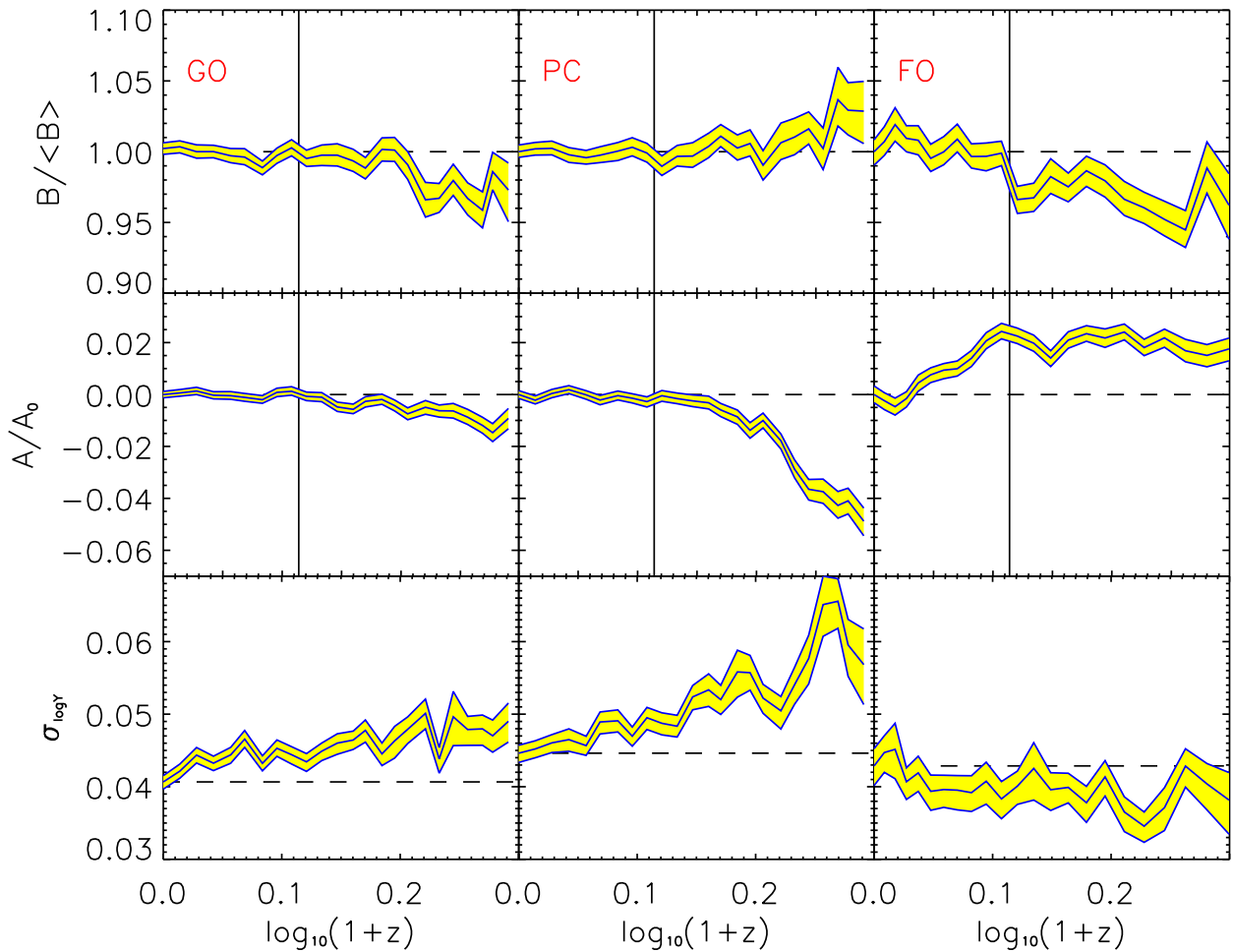}
\caption{
The dependence of the slope, normalisation and scatter of the $E(z)^{-2/3}Y_{500}-M_{500}$ 
relation with redshift, for the GO (left), PC (middle) and FO (right) simulations. Results
are plotted from $z=0$ to $z=1$.
In the top panels, the best-fit slope at each redshift is normalised to the median 
slope for outputs at $z<0.3$ (shown as a vertical line).
The middle panels illustrate the redshift dependence of the normalisation
after the self-similar dependence has been taken out; the normalisation
is divided by the $z=0$ value in this case. 
In the bottom panels, the rms
scatter in $\log_{10}Y$, $\sigma_{\log_{10} Y}$, 
is shown as a function of redshift. For both the 
normalisation and scatter values, the slope was fixed to the $z<0.3$ 
median value when performing the fits. 
The bands in all panels illustrate 16 and 84 per centiles, 
calculated by bootstrap resampling the data.
All three models predict very little evolution in the slope and normalisation of the 
$E(z)^{-2/3}Y_{500}-M_{500}$ relation to $z=1$ and the intrinsic scatter remains
small ($\sigma_{\log_{10} Y}<0.06$).
}
\label{fig:ymrel_evol}
\end{figure*}

We have also investigated the dependence of the $Y_{500}-M_{500}$ relation on redshift.
In the left-hand panels of Fig.~\ref{fig:ymrelz10}, we present results for $z=1$, allowing
a simple comparison to be made with the $z=0$ results for each model. It is evident that
the clusters evolve close to the self-similar expectation in all three models, given that
the normalisation and slope changes very little between the two redshifts 
(see also Table~\ref{tab:ymrelz0}). To quantify this further, we have also plotted the best-fit
normalisation, slope and scatter as a function of redshift in Fig.~\ref{fig:ymrel_evol},
where we used all available outputs from $z=0$ to $z=1$. (Equivalent plots 
for the other scaling relations are provided in the Appendix.)

The dependence of the best-fit slope with redshift for all three models is 
shown in the top panels of Fig.~\ref{fig:ymrel_evol}. For clarity we normalise the slope to the
median value at $z<0.3$ and the yellow bands indicate the uncertainties (using the
16/84 per centile values). All three models are consistent with no evolution in slope
to $z=0.3$, then some mild evolution is seen at higher redshift, where the number of
massive clusters drops. This evolution is very minor, however, as the slope remains 
within around 5 per cent of the low redshift value.

The variation in normalisation with redshift is shown in the middle panels of Fig.~\ref{fig:ymrel_evol}.
Here, we have fixed the slope at the $z<0.3$ median value and just allowed the single 
normalisation parameter to vary. Again, we factored out the self-similar evolution and normalised
to the $z=0$ result, so a value consistent with zero corresponds to self-similar evolution. 
In the GO and PC cases, the normalisation is consistent with self-similar evolution to $z=0.3$,
afterwards there is some negative evolution (i.e. the relation evolves slightly more slowly than
predicted from the self-similar model), especially in the PC case. The FO model shows different
behaviour: at low redshift ($z<0.3$), $Y_{500}$ increases more rapidly with redshift 
than the self-similar case (at fixed mass), then at higher redshifts evolves in accordance with
the self-similar expectation. These differences in evolution
are likely to be real and reflect the varying gas physics. In the GO case, the gas at high redshift is
slightly colder than expected (due to an increase in the merger/accretion rate leading to a larger
residual unthermalised component). In the PC case, the high redshift pre-heating leads to a deficit
in gas mass but the clusters start to recover at low redshift as the entropy scale at fixed mass
set by gravitation is larger. Finally, in the FO case, the feedback from black holes is stronger
at late times, leading to a decrease in gas content \citep{Short10}. 
In all three cases, however, the effect on the normalisation is still small; the largest change 
is from the PC model at $z=1$ where only a 10 per cent decrease is seen.

Finally, we illustrate how the scatter in the $Y_{500}-M_{500}$ relation 
evolves with redshift in the bottom panels of Fig.~\ref{fig:ymrel_evol}.
The $z=0$ value is also shown as a dashed horizontal line for clarity. Again, the picture is consistent with
minimal change; the scatter only increases to $z=1$ by 0.01 or so in the GO and PC cases, and decreases by less
than 0.01 in the FO case.

\subsection{Relationship between $Y_{500}$ and observables}
\label{sec:yobs}

\begin{figure}
\centering
\includegraphics[width=8.5cm]{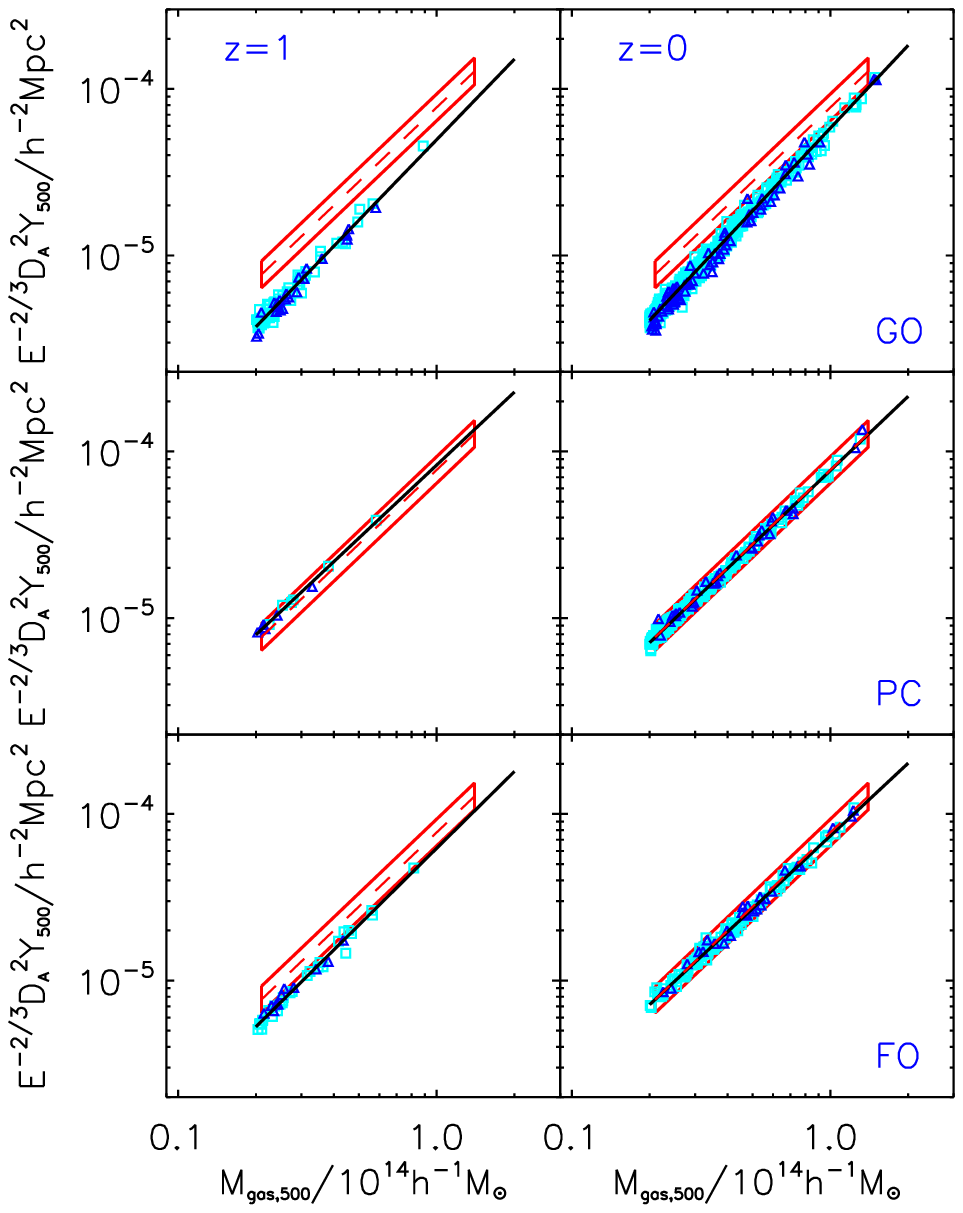}
\caption{
Scaling relations between $Y_{500}$ and hot gas mass, $M_{\rm gas,500}$
for the GO, PC and FO models at $z=1$ and $z=0$. Details for each panel
are the same as described in Fig.~\ref{fig:ymrelz10}. The scatter is particularly
small for the simulated version of this relation 
as the quantities plotted on the two axes are strongly correlated.}
\label{fig:ymgasrelz10}
\end{figure}

We also present scaling relations between $Y_{500}$ and other key X-ray observables.
Fig.~\ref{fig:ymgasrelz10} shows $Y_{500}-M_{\rm gas,500}$ relations, laid out as before. 
This relation is interesting to study because it essentially probes 
non self-similar
behaviour in the mass-weighted temperature, $T_{\rm m}$, of the gas, 
since $Y \propto M_{\rm gas}T_{\rm m}$ and thus $M_{\rm gas}$ appears on both axes.
Here we fit data within the range, $2\times 10^{13}\hMsol < M_{\rm gas,500} < 2\times 10^{14}\hMsol$.
As with the $Y_{500}-M_{500}$ relation, the slope from the GO model at $z=0$ is close to the
self-similar value of 5/3. The PC and FO models have shallower slopes, due
to the increase in the temperature of the gas in low-mass clusters. As might
be expected, the scatter in the relation is even tighter than for the $Y_{500}-M_{500}$
relation, and is typically 0.02-0.03. 
The distribution of the scatter is also close to log-normal.
From comparing the $z=1$ and $z=0$ results,
both GO and PC models predict evolution that is close to self-similar (the normalisation
is within 5 per cent of the $z=0$ value out to $z=1$)  but the FO relation
evolves more slowly with redshift ($\sim 10$ per cent lower at $z=1$), 
again due to the increase in feedback from the AGN at late times that additionally heats the gas.
This evolutionary behaviour is confirmed when studying the relation at all available redshifts from
$z=0$ to $z=1$, in Fig.~\ref{fig:ymgasrel_evol}, which also shows that the slope and scatter vary little.

\begin{figure}
\centering
\includegraphics[width=8.5cm]{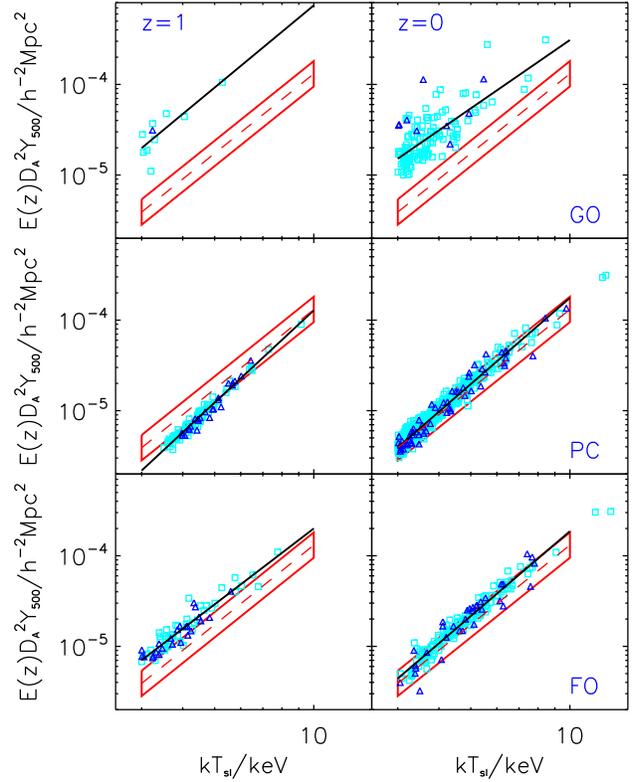}
\caption{
Scaling relations between $Y_{500}$ and X-ray 
spectroscopic-like temperature, $T_{\rm sl}$, 
evaluated outside the core. Details for each panel
are the same as described in Fig.~\ref{fig:ymrelz10}.
This relation has relatively large scatter for both
the observations and simulations, since $T_{\rm sl}$
is a noisier property than the other observables being considered in this study.}
\label{fig:ytrelz10}
\end{figure}

We also consider scaling relations between $Y_{500}$ and X-ray spectroscopic-like temperature,
$T_{\rm sl}$, and show results in Fig.~\ref{fig:ytrelz10}, with the redshift dependence of the slope,
normalisation and scatter illustrated in Fig.~\ref{fig:ytrel_evol}. 
Here, we further restrict our sample to contain only clusters with $kT_{\rm sl}>3 \, {\rm keV}$, as
the spectroscopic-like temperature only applies to hot clusters where thermal bremsstrahlung 
dominates the X-ray emission. This reduces our samples to 136 (12), 583 (102) and 
179 (73) clusters at $z=0$ ($z=1$) in the GO, PC and FO models respectively. Note the more
severe reduction in the GO case; the non-gravitational heating in the PC and FO models
increases $T_{\rm sl}$ at fixed mass, relative to the GO case, and thus increases the number
of clusters in their X-ray temperature-limited samples. 
Best-fit relations are then calculated for clusters in the range, 
$3 \, {\rm keV} < kT_{\rm sl} < 10 \, {\rm keV}$. 

The GO model relation has a slope that is consistent with the self-similar expectation
($B=5/2$) at $z=0$ and $z=1$. The relation evolves slightly faster than the self-similar
model (the normalisation is around 10 per cent higher than expected at $z=1$), while the
scatter is approximately constant at all redshifts, but is much higher 
than for the previous relations ($\sigma_{\log_{10}Y} \simeq 0.15-0.2$). This last point is
due to $T_{\rm sl}$ being a much noiser property as it is sensitive to the clumpy, low
entropy gas that is prevalent in this model. 
We also note that the scatter 
is poorly described by a log-normal distribution.
In comparison, the PC and FO models, which
have much smoother gas, typically have lower scatter, $\sigma_{\log_{10}Y} \simeq 0.05-0.1$, 
that is well described by a log-normal distribution.
The slope in these two models is significantly steeper ($B\simeq 3$) and the evolution 
of this relation shows the largest departure from self-similarity (up to 20 per cent
lower/higher at $z=1$ in the PC/FO models).

\begin{figure}
\centering
\includegraphics[width=8.5cm]{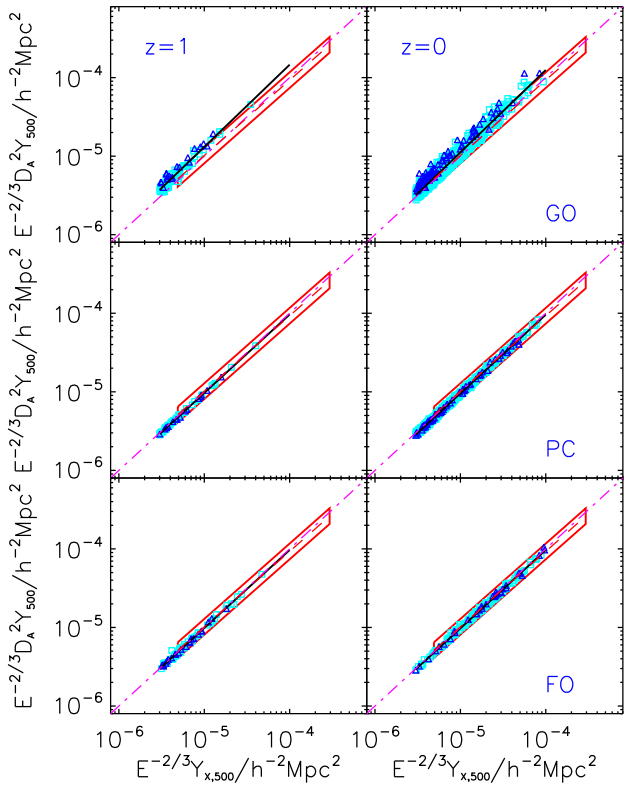}
\caption{Scaling relations between the SZ $Y_{500}$ and the X-ray analogue,
$Y_{\rm X,500}$. Details for each panel are the same as described in 
Fig.~\ref{fig:ymrelz10}. The dot-dashed line corresponds to 
$Y_{500}=Y_{\rm X,500}$; only the GO simulation shows
a significant offset from this relation, due to the presence of 
clumpy low-entropy gas.}
\label{fig:yyxrelz}
\end{figure}

Finally, in Fig.~\ref{fig:yyxrelz} we plot $Y_{500}$ against $Y_{\rm X,500}$ 
for our cluster samples and show the redshift dependence of the scaling relation
parameters in Fig.~\ref{fig:yyxrel_evol}. We do this to directly highlight 
how the choice of gas temperature affects the results: any deviation from 
$Y_{500}=Y_{\rm X,500}$ must be due to differences between mass and X-ray weighted 
temperatures. No significant deviation is seen in the PC and FO models 
(the difference in normalisations at $E(z)^{-2/3}Y_{X,500}=10^{-5}\YSZunits$ 
is less than 5 per cent) and there is very little scatter ($\sigma_{\log_{10}Y} \simeq 0.01-0.02$) 
at low and high redshift, that again has a distribution that is log-normal. 
The GO model, on the other hand, shows
a significant bias, such that $Y_{500} \simeq 1.1 Y_{\rm X,500}$
at $z=0$, increasing to $Y_{500} \simeq 1.3 Y_{\rm X,500}$ at $z=1$. The scatter is also 
significantly larger than for the other two models, $\sigma_{\log_{10}Y} \simeq 0.05$, and the distribution is skewed to lower values. 
Again, these results demonstrate that the clumpier gas in the GO model has a stronger
effect on the X-ray properties than the SZ properties. As we shall see in Section~\ref{sec:hse}, this 
has important consequences for our hydrostatic mass estimates.

\subsection{Effect of dynamical state}

It is also interesting to consider whether clusters undergoing mergers
are offset from the main $Y_{500}$ scaling relations as they could add to
the intrinsic scatter. We mark our {\it disturbed} ($s>0.1$) 
sub-samples as triangles in each of the figures presenting scaling relations, discussed
above (Figs.~\ref{fig:ymrelz10}-\ref{fig:yyxrelz}). Note that while a large value of $s$
is indicative of an ongoing merger, not all dynamically disturbed clusters have 
large values of $s$ \citep{Rowley04}.

As predicted from studying the hot gas pressure profiles in Section~\ref{sec:pressureprofiles},
the only significant offset seen between regular and disturbed clusters is for
the GO model, where disturbed objects lie slightly below the $Y_{500}-M_{500}$
and $Y_{500}-M_{\rm gas,500}$ relations, and above the $Y_{500}-Y_{\rm X,500}$
relation (there are not enough disturbed clusters to say anything conclusive
for the $Y_{500}-T_{\rm sl}$ relation). This suggests that there is a significant
difference in the fraction of unthermalised energy between regular and disturbed
clusters in this model. In the case of the $Y_{500}-M_{500}$ and $Y_{500}-M_{\rm gas,500}$
relations, the mass-weighted temperature is lower for disturbed clusters of the same
mass than regular clusters, leading to the negative offset. The effect is exacerbated 
when $T_{\rm sl}$ is considered (since it is weighted towards the cooler component),
leading to a positive offset in the $Y_{500}-Y_{\rm X,500}$ relation. 

\subsection{Comparison of $Y_{500}-M_{500}$ relation from other simulations}
\label{subsec:ymsim}

Given the importance of the $Y_{500}-M_{500}$ relation for cosmological applications
and its apparent insensitivity to cluster gas physics, it is important to 
compare our results to those from other groups using different simulations. 
A number of studies have been performed with varying assumptions for both the cosmology 
and gas physics, as well as the use of different algorithms for the $N$-body and
hydrodynamics solvers
(e.g. \citealt{White02,daSilva04,Motl05,Nagai06,Bonaldi07,Hallman07,Aghanim09,Yang10,Krause12,Battaglia11}).

\begin{figure}
\centering
\includegraphics[width=8.5cm]{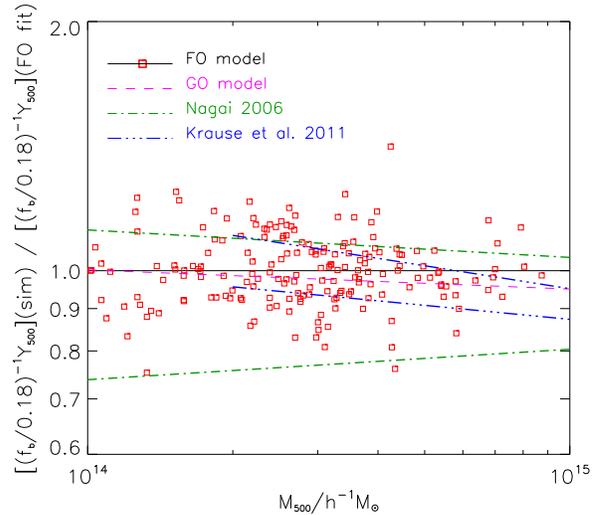}
\caption{Comparison between the FO $Y_{500}-M_{500}$ scaling relation 
and results from other simulations at $z=0$. All results are normalised to
the best-fit FO $Y_{500}-M_{500}$ relation and $Y_{500}$ values from other simulations
have been re-scaled to account for differences in baryon fraction (see text for details). 
The square symbols are individual cluster $Y_{500}$ values from the FO model and the 
dashed line the best-fit GO relation. The dot-dashed lines are best-fit relations
from \citet{Nagai06}; the upper line corresponds to their non-radiative (AD) simulation
and the lower line their run with cooling and star formation (CSF). Finally, the 
triple-dot-dashed lines are taken from \citet{Krause12} for their restricted A and B 
samples (upper and lower lines respectively).} 
\label{fig:ymrel_comp_sims}
\end{figure}

We choose to compare our results with the work of \citet{Nagai06} and \citet{Krause12} for
two reasons. Firstly, both groups presented results for  $\Delta=500$ and are thus most 
readily comparable with ours. Secondly, the two groups used very different codes, so it is
useful to also include that uncertainty in our comparison.
In Fig.~\ref{fig:ymrel_comp_sims} we compare our best-fit $Y_{500}-M_{500}$ relation 
at $z=0$ from the FO model (solid line) with the results of these authors. To highlight
the differences between simulations, we normalise all results to our best-fit FO relation.
We also have to make a correction for the different baryon fractions used in the simulations,
since $Y \propto f_{\rm b}$. In both cases, the baryon
fraction is lower than our adopted value of $f_{\rm b}=0.18$ 
(\citealt{Nagai06} assumed $f_{\rm b}=0.14$ and 
\citealt{Krause12} assumed $f_{\rm b}=0.133$). Note that this is not a perfect correction
as it does not account for the non-self-similar behaviour of the baryon fraction with cluster
mass.

\citet{Nagai06} presented results for 11 clusters simulated with the {\sc art} code
(e.g. \citealt{Kravtsov02}),
that uses the adaptive mesh refinement technique to model hydrodynamics. Two sets of
runs were studied, a non-radiative run (labelled AD) and a run with cooling and star
formation (labelled CSF). Out of the 11 clusters, 6 have $M_{500}>10^{14}\hMsol$ 
(c.f. our FO model with 188 clusters in this mass range). The upper dot-dashed line in 
Fig.~\ref{fig:ymrel_comp_sims} is their best-fit relation to the AD clusters. The 
slope of their relation ($1.66 \pm 0.09$)  is in agreement with our (non-radiative) GO result 
($1.670 \pm 0.007$; dashed line) while the normalisation is within
10 per cent of ours. Such good agreement is reassuring given the different hydrodynamic
methods used, although the large difference in sample size must be borne in mind.
Their CSF result is shown as the lower dot-dashed curve in Fig.~\ref{fig:ymrel_comp_sims}; comparing
with our FO relation, their normalisation is significantly lower (20-30 per cent). As the 
author points out, the reduction in SZ signal in the CSF run is mainly due to the lower gas 
fraction caused by (over-)cooling and star formation that removes hot gas from the ICM. As we discussed 
earlier, the gas fractions in the FO run are also lower than in the non-radiative case but the mechanism 
responsible (strong feedback) compensates for this by heating the gas to a higher temperature. 

\citet{Krause12} present results for two cluster samples, A and B, shown as the upper and lower
triple-dot-dashed lines in the figure. Both samples were simulated with the same {\sc gadget2} 
$N$-body/SPH code as used in this study but contained different assumptions for the gas physics. Sample A
contained 39 clusters re-simulated from a large parent volume while sample B was a mass-limited
sample of 117 objects, taken from a single simulation. While both samples are larger than in
\citet{Nagai06} the number of massive clusters is still significantly smaller than in our FO sample.  
The two samples (we show results restricted to clusters with $M_{500}>2\times 10^{14}\hMsol$) 
compare well with ours once the different baryon fraction is scaled out. The normalisation in
both cases is within 10 per cent or so, although the slope is slightly flatter, a result that
appears only marginally significant (the slope of sample B is $1.64 \pm 0.03$ compared with the FO slope of
$1.69 \pm 0.02$).

\subsection{Comparison with observational data}

We have also compared our results to observational data, now that blind SZ surveys are starting
to yield significant numbers of (SZ-selected) clusters, enabling estimates of the 
$Y_{500}-M_{500}$ relation to be performed \citep{Andersson11,Planck11c}. Here, we compare
our results with those from the {\it Planck} Collaboration 
(\citealt{Planck11c}, hereafter PXMM), although we note that their 
best-fit $Y_{500}-M_{500}$ relation is similar to the SPT result derived from a lower number
of clusters by \citet{Andersson11}. 

The PXMM sample consists of 62 clusters with $z<0.5$ and used X-ray data from {\it XMM-Newton} to
define the size ($r_{500}$) and mass ($M_{500}$) of each cluster, calibrated using the X-ray 
$M_{500}-Y_{\rm X,500}$ relation previously derived by \citet{Arnaud10}. 
Once the cluster size was defined, the SZ flux was measured using a 
multi-frequency matched-filter technique, based
on the ICM pressure profile of \citet{Arnaud10}. We show their best-fit results to the $Y_{500}-M_{500}$,
$Y_{500}-M_{\rm gas,500}$ and $Y_{500}-T_{\rm sl}$ relations as dashed lines and
illustrate their intrinsic scatter with boxes, in 
Figs.~\ref{fig:ymrelz10}-\ref{fig:ytrelz10}. (Note we show these in both panels to help
gauge the sense of evolution in our simulated relations, but the observed fits are more applicable
to our $z=0$ results.)

It is remarkable how well the PXMM results agree with our PC and FO models; only the 
$Y_{500}-T_{\rm sl}$ relation shows any obvious offset but that is nevertheless small. The 
reason for such good agreement is not obvious or necessarily expected, given the complicated
procedure involved in deriving the observed parameters (we are using the simplest form of the
simulated $Y_{500}-M_{500}$ relation here). 

Another interesting result from the PXMM sample is that the results are consistent with 
$Y_{500}=Y_{\rm X,500}$ on average (again like our PC and FO models), however the scatter in the
observed relation is significantly larger than ours (observationally, $\sigma_{\log_{10}Y}=0.1$, 
around a factor of 5 larger than for our PC and FO simulations). As a result, the scatter in the
other observed PXMM scaling relations are also larger than ours; e.g. the scatter is 2-3
times larger for the $Y_{500}-M_{500}$ relation. Thus if our PC and FO simulations, calibrated to X-ray data, 
provide faithful estimates of the mean SZ/X-ray scaling relations, observational estimates of
the quantities must somehow increase the scatter without introducing significant bias. One potential
source of scatter is due to the projection of large-scale structure along the line-of-sight; we investigate
this below. 

\subsection{Projection effects}
\label{subsec:projection}

\begin{table*}
\caption{Best-fit parameters for simulated SZ scaling relations using
projected (cylinder) values from cluster and sky maps (see text for 
further details).
Column 1 gives the scaling relation being considered; 
column 2 the redshift range;
column 3 the simulation model;
column 4 the number of clusters used in the fit;
columns 5 \& 6 list the best-fit values for the normalisation and slope parameters respectively;
and column 7 lists the estimated scatter in $\sigma_{\log_{10}Y}$. 
Quoted uncertainties correspond to either the 16th or 84th per centile 
(whichever is largest), estimated using the bootstrap re-sampling technique.}

\begin{center}
\begin{tabular}{lllcccc}
\hline
Flux & Redshift & Model & $N_{\rm clus}$ & $A$ & $B$ & $\sigma_{\log_{10}Y}$\\
\hline
$Y_{500}^{\rm clus}-M_{500}$ 
& $0<z<0.5$ & GO & 1346 & $-4.677 \pm 0.003$ & $1.650 \pm 0.007$ & $0.045 \pm 0.001$\\
&           & PC & 1074 & $-4.671 \pm 0.004$ & $1.72 \pm 0.01$   & $0.068 \pm 0.002$\\
& $0.5<z<1$ & GO & 2952 & $-4.702 \pm 0.002$ & $1.613 \pm 0.007$ & $0.050 \pm 0.001$\\
&           & PC & 2199 & $-4.699 \pm 0.003$ & $1.74 \pm 0.01$   & $0.077 \pm 0.001$\\
\hline
$Y_{500}^{\rm sky}-M_{500}$ 
& $0<z<0.5$ & GO & 1346 & $-4.622 \pm 0.003$ & $1.507 \pm 0.009$ & $0.059 \pm 0.001$\\
&           & PC & 1074 & $-4.455 \pm 0.003$ & $1.23  \pm 0.01$  & $0.059 \pm 0.001$\\ 
& $0.5<z<1$ & GO & 2952 & $-4.677 \pm 0.003$ & $1.524 \pm 0.007$ & $0.062 \pm 0.001$\\
&           & PC & 2199 & $-4.556 \pm 0.003$ & $1.32  \pm 0.02$  & $0.071 \pm 0.001$\\ 
\hline
$Y_{500}^{\rm skysub}-M_{500}$ 
& $0<z<0.5$ & GO & 1346 & $-4.686 \pm 0.004$ & $1.71 \pm 0.01$ & $0.074 \pm 0.003$\\
&           & PC & 1074 & $-4.683 \pm 0.005$ & $1.80 \pm 0.02$ & $0.106 \pm 0.003$\\
& $0.5<z<1$ & GO & 2952 & $-4.719 \pm 0.003$ & $1.666 \pm 0.009$ & $0.071 \pm 0.001$\\
&           & PC & 2199 & $-4.721 \pm 0.004$ & $1.80 \pm 0.01$ & $0.104 \pm 0.002$\\
\hline
\end{tabular}
\end{center}
\label{tab:ymrelcyl}
\end{table*}

%The low scatter in our $Y_{500}(Y_{\rm X})-M_{500}(Y_{\rm X})$ relation, 
%as well as the large scatter in the observed (PXMM) $Y_{500}-Y_{\rm X,500}$ relation,
%suggests that the scatter in the PXMM $Y_{500}-M_{500}$ relation is entirely
%due to effects associated with the observed estimation of $Y_{500}$. %
%Many factors could contribute to this and a thorough investigation would 
%ultimately require us to run our simulated clusters through a full observational pipeline
%(including the use of multi-frequency matched filters in the case of the {\it Planck}
%data). Here, we concentrate on one other  obvious {\it physical} origin for the scatter: the 
%contribution to the SZ signal from additional hot gas along the line-of-sight. We can also
%check whether such projection effects also introduce any bias to the $Y-M$ relation.

As detailed in Section~\ref{sec:skymaps} we have constructed 50 $5^{\circ} \times 5^{\circ}$  
mock realisations of the SZ sky (Compton $y$ maps) from our GO and PC simulations. 
(Unfortunately, it is not currently possible to do this for the FO model as it was not run
on the full Millennium volume.) We use these maps to estimate the (cylindrical) $Y_{500}$
for the clusters that are present, as follows. 

Firstly, we cross-match our 50 maps with cluster catalogues at all available redshifts (catalogues are
constructed for all snapshots used to make the maps, providing there are objects above
our mass limit of $M_{500}=10^{14}\hMsol$). This is done by performing the same 
operations (translation, rotation, reflection) on the cluster centre co-ordinates as 
was done with each of the snapshots, then finding the pixel in the map that corresponds
with the cluster centre, for those objects within the map region. We then identify which
pixels fall within the projected radius, $R_{500}=r_{500}$, and compute the SZ $Y_{500}$ value
which we define as
\begin{equation}
Y_{500}^{\rm sky} = D_{\rm A}^2 \, \delta \Omega \, \sum_{i,j} \, y_{i,j},
\label{eqn:ysky}
\end{equation}
where the sum is performed over all relevant pixels (with indices, $i,j$) and
$\delta \Omega$ is the solid angle of each pixel (we use $1200\times 1200$ pixels
so assume $\delta \Omega = 0.25 \times 0.25 \, {\rm arcmin}^2$). Finally, we throw away clusters
that have a more massive neighbour whose centre lies within its own radius, $R_{500}$, as this
interloper would dominate the estimated SZ flux. Our final catalogue is restricted to 
clusters with $M_{500}>10^{14}\hMsol$ and $z<1$; for comparative purposes we split this
into a low-redshift ($z<0.5$) and high-redshift ($z>0.5$) sub-samples. The number of clusters 
in each of these sub-samples for the GO and PC models are listed in Table~\ref{tab:ymrelcyl}.
The larger numbers in the high-redshift sample are expected due to the larger volume there 
(for fixed solid angle). Note that the same cluster could appear more than once (in a different
realisation or redshift). 

In order to extract the cluster signal from the rest of the large-scale structure
along the line-of-sight, we also compute cylindrical $Y_{500}$ values due to the cluster
region itself. To do this, we apply equation~(\ref{eqn:ysky}) to our cluster maps, 
detailed in Section~\ref{sec:clustermaps}. As a reminder, the length of the cylinder,
centred on the cluster, is $z=12r_{500}$; this approximately corresponds to three virial
radii from the centre in each direction along the line-of-sight. We refer to this
$Y$ value as $Y_{500}^{\rm clus}$; clearly $Y_{500}^{\rm sky}>Y_{500}^{\rm clus}$ by
definition.

\begin{figure}
\centering
\includegraphics[width=8.5cm]{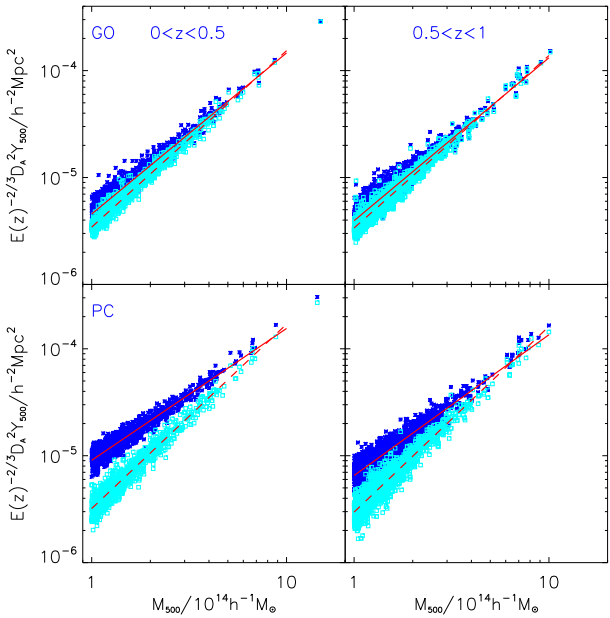}
\caption{Projected $Y_{500}-M_{500}$ relations for clusters in the GO
(top panels) and PC (bottom panels) sky maps with $0<z<0.5$ (left panels)
and $0.5<z<1$ (right panels). The stars correspond to $Y_{500}^{\rm sky}$
values, i.e. calculated from the full sky map. The squares correspond to 
$Y_{500}^{\rm clus}$ values, i.e. from the cluster region. In both cases the
true $M_{500}$ values were used. The dashed line is a best-fit
to the $Y_{500}^{\rm clus}-M_{500}$ relation and the solid line to the
$Y_{500}^{\rm sky}-M_{500}$ relation. The $Y_{500}^{\rm sky}$ values
are higher on average than $Y_{500}^{\rm clus}$, especially in the PC
simulation at low mass and low redshift, where the difference is a factor of
2-3.}
\label{fig:y500m500cyl}
\end{figure}

The squares in Fig.~\ref{fig:y500m500cyl} represent the $Y_{500}^{\rm clus}-M_{500}$
relation for our GO (top panels) and PC (bottom panels) models at high (left panels)
and low (right panels) redshifts. We re-scale cluster $Y_{500}^{\rm clus}$ values
by $E(z)^{-2/3}$ to account for evolution across the redshift range in each panel.
Best-fit parameters ($A,B,\sigma_{\rm log Y}$) are given in Table~\ref{tab:ymrelcyl};
a pivot mass of $3\times 10^{14}\hMsol$ was adopted for all the fits.
The GO model relations show similar trends to those seen in the spherical 
$Y_{500}-M_{500}$ relation; the slope is close to self-similar and the scatter is small.
The PC relations again have slopes that are steeper than the self-similar value 
but also have slightly larger scatter ($\sigma_{\log_{10}Y}\simeq 0.07-0.08$), reflecting
in part the effect of additional evolution with redshift.

The stars in Fig.~\ref{fig:y500m500cyl} are for when $Y_{500}^{\rm sky}$ values are 
used and thus contain the additional signal from beyond the cluster. The difference
between the two relations in each panel (as can be seen from the best-fit lines) 
is most prominent for the PC model, where the slope has decreased from $\sim 1.7$
to $\sim 1.2$, due to $Y_{500}^{\rm sky}$ being significantly larger than 
$Y_{500}^{\rm clus}$ in the lower mass objects. As was discussed in 
Section~\ref{sec:simdetails}, the pre-heating was applied everywhere at $z=4$ and 
thus substantially increased the thermal energy of the gas, as indicated by the 
three-fold increase in the mean $y$ signal. Such widespread heating is likely 
to be unrealistic as it would require a huge amount of energy and would boil off 
the small amount of neutral hydrogen and helium in the IGM
\citep{Theuns01,Borgani09}, so the PC result represents a worse-case scenario for the effects
of projection on the $Y$ signal.

\begin{figure}
\centering
\includegraphics[width=8.5cm]{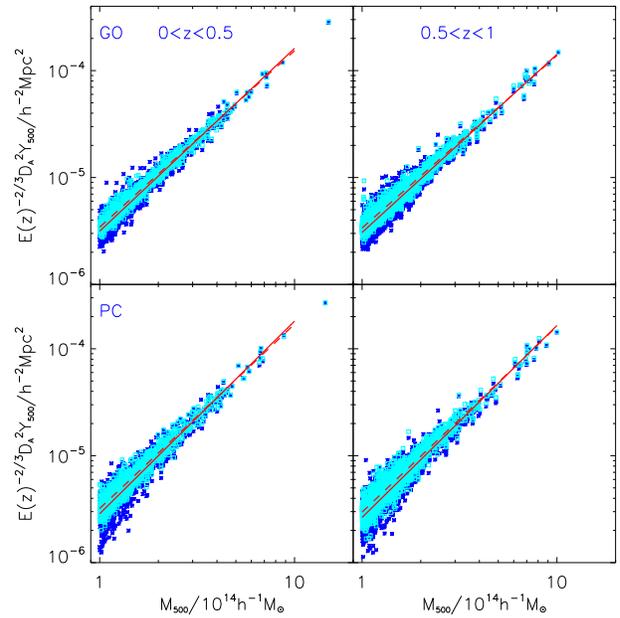}
\caption{As in Fig.~\ref{fig:y500m500cyl} but the $Y_{500}^{\rm sky}$
values have had the mean background signal subtracted. The two best-fit 
relations are now very similar in all panels.}
\label{fig:y500m500cylbsub}
\end{figure}

Observations of the SZ effect made with the {\it Planck} satellite are unable 
to measure the mean $y$ signal as at each frequency, spatial temperature
fluctuations are measured with respect to the all-sky mean.
It is therefore more realistic to compare the 
{\it background-subtracted} values of $Y_{500}^{\rm sky}$ to the cluster values. To
do this we compute the projected angular area for each cluster and compute the 
expected contribution to $Y_{500}$ from the mean $y$
\begin{equation}
Y_{500}^{\rm skysub}=Y_{500}^{\rm sky} - \left< y \right> \, D_{\rm A}^2 \, \Omega_{500},
\label{eqn:Y500skysub}
\end{equation} 
where $\Omega_{500}$ is the solid angle subtended by the cluster out to a projected
radius, $R_{500}$. The results of this procedure are shown in Fig.~\ref{fig:y500m500cylbsub},
with best-fit parameters for the $Y_{500}^{\rm skysub}-M_{500}$ relations given in 
Table~\ref{tab:ymrelcyl}. 

Interestingly, the two best-fit relations are now almost identical for each 
run and within each redshift range. A simple background subtraction therefore 
removes any bias in the mean relation generated from the additional hot gas
along the line-of-sight. The scatter is considerably larger in the 
$Y_{500}^{\rm skysub}-M_{500}$ relation, in part due to the fact that the additional 
signal is not constant everywhere. The PC relations again contain the largest scatter,
comparable to the observed scatter in the PXMM data ($\sigma_{\log_{10}Y} \simeq 0.1$). 
Although the result is model dependent, it is clear that part (if not all) of the observed
scatter can be attributed to projection effects.
%\vspace{-0.5cm}

\section{Hydrostatic bias}
\label{sec:hse}
%\subsection{Estimated scaling relations assuming hydrostatic equilibrium}

\begin{table*}
\caption{Best-fit parameters for simulated  SZ scaling relations, with estimated
properties from the assumption of hydrostatic equilibrium.
Column 1 gives the scaling relation being considered; 
column 2 the redshift;
column 3 the simulation model;
column 4 the number of clusters used in the fit;
columns 5 \& 6 list the best-fit values for the normalisation and slope parameters respectively;
and column 7 lists the estimated scatter in $\sigma_{\log_{10}Y}$ or $\sigma_{\log_{10} M}$ 
(whichever is appropriate). 
Quoted uncertainties correspond to either the 16th or 84th per centile 
(whichever is largest), estimated using the bootstrap re-sampling technique.}
\begin{center}
\begin{tabular}{lllcccc}
\hline
Relation & Redshift & Model & $N_{\rm clus}$ & $A$ & $B$ & $\sigma_{\log_{10}Y/M}$\\
\hline
$E(z)^{-2/3} Y_{500}^{\rm HSE}-M_{500}^{\rm HSE}$ 
& $z=0$ & GO & 439 & $-4.54 \pm 0.02$  & $1.60 \pm 0.07$ & $0.186 \pm 0.008$ \\
&       & PC & 738 & $-4.75 \pm 0.009$ & $1.69 \pm 0.03$ & $0.111 \pm 0.005$ \\
&       & FO & 179 & $-4.69 \pm 0.01$  & $1.50 \pm 0.05$ & $0.13 \pm 0.01$ \\
\cline{3-7}
& $z=1$ & GO & 25  & $-4.6  \pm 0.3$   & $1.6  \pm 0.8$  & $0.21 \pm 0.03$ \\
&       & PC & 94  & $-4.96 \pm 0.04$  & $1.5  \pm 0.1$  & $0.11 \pm 0.01$ \\
&       & FO & 57  & $-4.76 \pm 0.06$  & $1.4  \pm 0.2$  & $0.14 \pm 0.02$ \\
\hline
$E(z)^{2/5} M_{500}^{\rm HSE}-Y_{X,500}^{\rm HSE}$ 
& $z=0$ & GO & 787 & $0.44 \pm 0.01$ & $0.58 \pm 0.02$ & $0.118 \pm 0.004$ \\
&       & PC & 672 & $0.571 \pm 0.007$ & $0.545 \pm 0.009$ & $0.064 \pm 0.003$ \\
&       & FO & 179 & $0.57 \pm 0.01$   & $0.60 \pm 0.02$ & $0.080 \pm 0.007$ \\
\cline{3-7}
& $z=1$ & GO & 98  & $0.37 \pm 0.06$ & $0.50 \pm 0.07$ & $0.13 \pm 0.01$ \\
&       & PC & 86  & $0.69 \pm 0.03$ & $0.57 \pm 0.04$ & $0.066 \pm 0.008$ \\
&       & FO & 74  & $0.56 \pm 0.02$ & $0.59 \pm 0.04$ & $0.09 \pm 0.01$ \\
\hline
$E(z)^{-2/3} Y_{500}^{Y_{\rm X}}-M_{500}^{Y_{\rm X}}$
& $z=0$ & GO & 398 & $-4.356 \pm 0.009$ & $1.84 \pm 0.02$ & $0.055 \pm 0.002$ \\
&       & PC & 736 & $-4.694 \pm 0.001$ & $1.840 \pm 0.004$ & $0.0150 \pm 0.0005$ \\
&       & FO & 175 & $-4.668 \pm 0.002$ & $1.659 \pm 0.008$ & $0.019 \pm 0.001$ \\
\cline{3-7}
& $z=1$ & GO & 31  & $-4.29 \pm 0.03$ & $2.09 \pm 0.09$ & $0.052 \pm 0.007$ \\
&       & PC & 102 & $-5.040 \pm 0.002$ & $1.757 \pm 0.005$ & $0.0092 \pm 0.0008$ \\
&       & FO & 75  & $-4.799 \pm 0.004$ & $1.67 \pm 0.02$ & $0.022 \pm 0.006$ \\
\hline
\end{tabular}
\end{center}
\label{tab:ymrelhse}
\end{table*}

In the previous section, we saw that our PC and FO models produced SZ/X-ray scaling
relations that were in good agreement with the PXMM observational data. A significant
uncertainty in the observational determination of scaling relations is the 
(direct or indirect) assumption of hydrostatic equilibrium (HSE), required for deriving the 
cluster mass ($M_{500}$) and radius ($r_{500}$). It is therefore interesting to look
at the accuracy of this assumption in our simulations as the good agreement between our
results and the observations can only be preserved if hydrostatic bias is small (in the
absence of additional systematic effects). 

For a cluster in HSE, the pressure gradient in the ICM is sufficient to balance 
gravity; the total mass of the cluster can then be calculated as
\begin{equation}
M^{\rm HSE}(<r) = - \, {kT r \over G \mu m_{\rm H}} \, 
\left[
  {{\rm d}\ln \rho \over {\rm d}\ln r} + 
  {{\rm d}\ln T \over {\rm d}\ln r}
\right]
\label{eqn:hse}
\end{equation}
where $\mu=0.59$ is the mean molecular weight for an ionised plasma (assuming zero
metallicity). We use the spectroscopic-like temperature to evaluate the local temperature,
$T(r)$ and its gradient, $d\ln T / d\ln r$, at radius, $r$.

Estimation of the cluster mass based on hydrostatic equilibrium
can be biased for three reasons. Firstly, the estimated mass 
within a fixed radius can be different from the true mass 
because the intracluster gas is not perfectly hydrostatic. 
Previous simulations have shown that mass estimates can be 
too low by up to 20 per cent, due to incomplete thermalisation of the gas
(e.g. \citealt{Evrard96,Rasia04,Kay04,Rasia06,Kay07,Nagai07a,Nagai07b,
Piffaretti08,Ameglio09,Lau09}). A second effect is that the X-ray 
temperature of the gas may be lower than the mean (mass-weighted) temperature.
Such an effect depends on the thermal structure of the gas (in particular, the
low entropy tail associated with substructure) and can be particularly severe
when radiative cooling effects are strong. 
\footnote{We note that \citet{Nagai07a} 
found the X-ray temperature to be higher than the mass-weighted temeprature
in a mock {\it Chandra} analysis of their simulated clusters, but they exclude
any resolved cold clumps from their calculation.}
Finally, the cluster's size itself 
is usually defined as a scale radius (e.g. $r_{500}$) which is mass-dependent
so also depends on the assumption of hydrostatic equilibrium. 

To study how these combined effects impact upon our scaling relations, 
we estimate the hydrostatic mass of each cluster as follows. Firstly, 
we compute the hot gas ($T>10^{5}\,{\rm K}$) density and temperature
profiles. In lower mass clusters the profiles can get rather noisy
due to limited particle numbers which can affect the estimation of the pressure
gradient. To avoid this, we fit a cubic polynomial function to each profile 
(in log space) to generate a smoothed representation. (This also has the 
advantage that the gradient can be derived analytically.) We then use these
model profiles to estimate the mass, $M^{\rm HSE}$, using equation~(\ref{eqn:hse}), 
then vary the radius, $r$, until the following equation is satisfied
\begin{equation}
M_{500}^{\rm HSE} = {4 \pi \over 3} \, \left( r_{500}^{\rm HSE} \right)^3 \, 
500 \rho_{\rm cr}(z), 
\label{eqn:m500hse}
\end{equation}
where $M_{500}^{\rm HSE}$ and $r_{500}^{\rm HSE}$ are our estimated mass and
radius respectively. Once the radius is known we can use this to estimate 
the SZ flux which we will denote $Y_{500}^{\rm HSE}$. Again, this is the flux
from within a sphere centred on the cluster; all that has changed is the assumed
value of $r_{500}$. In what follows, we only consider the sub-set of clusters in 
the estimated mass range, $10^{14}\hMsol<M_{500}^{\rm HSE}<10^{15}\hMsol$. The 
numbers of clusters are listed for each model and redshift in Table~\ref{tab:ymrelhse}.

\subsection{Effect of hydrostatic assumption on cluster mass}

\begin{figure}
\centering
\includegraphics[width=8.5cm]{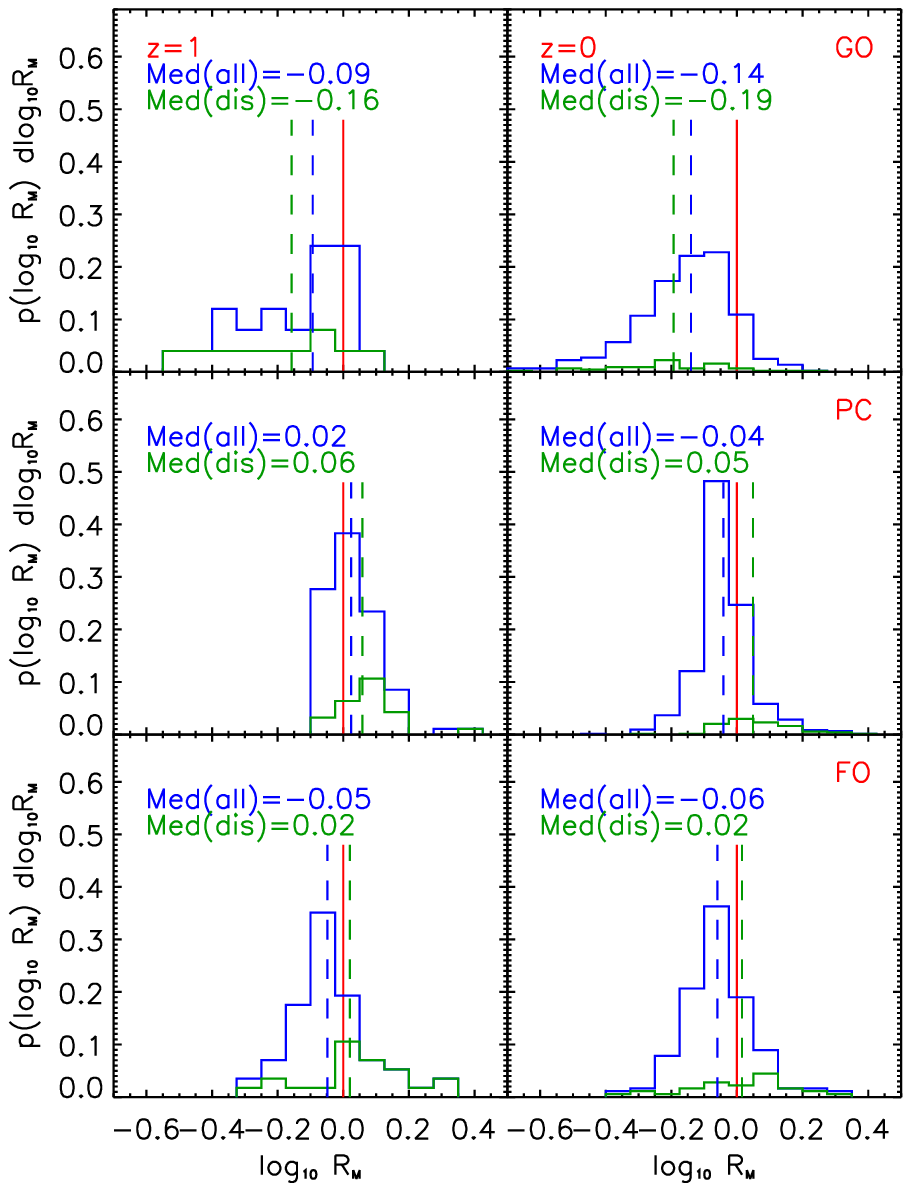}
\caption{Distribution of estimated-to-true mass ratios, $R_{\rm M}$, 
within the estimated $r_{500}$ for clusters at $z=1$ (left panels)
and $z=0$ (right panels). The top panels show results from 
the GO model, middle panels from PC and bottom panels from FO.
The green histogram is for the whole cluster sample while the 
blue histogram is for the disturbed sub-sample. Vertical dashed
lines indicate the median mass ratio for each case, with values
given in the legend. The median $R_{\rm M}$ is significantly smaller
in the runs with non-gravitational heating.}
\label{fig:mhseratio}
\end{figure}

We quantify the effect of the hydrostatic assumption on cluster mass
by considering the distribution of the estimated-to-true mass ratio, 
$R_{\rm M}\equiv \log_{10}(M_{500}^{\rm HSE}/M_{500})$, for our models
at $z=1$ and $z=0$. (Note that $R_{\rm M}$ directly measures the resulting 
shift along the logarithmic mass axis.) The results are shown in Fig.~\ref{fig:mhseratio}. 

At $z=0$, the GO results show a significant spread in mass ratios as
well as a large negative bias; the median value is $R_{\rm M}=-0.14$. In the
PC and FO models, the spread and bias is smaller, with the median increasing
to around $-0.05$. A similar situation is evident at $z=1$. The disturbed sub-sample,
where HSE should definitely not be a good approximation, shows a small offset 
in the median $R_{\rm M}$ from the overall sample; in the PC and FO cases the offset
is positive whereas in the GO case it is negative.

It is perhaps not surprising that the discrepancy between estimated and true
mass from the GO simulation is significantly higher than for the PC and FO models.
As is evident from the $Y_{500}-Y_{\rm X,500}$ relation (Fig.~\ref{fig:yyxrelz}), 
the former model predicts a more clumpy
intracluster medium due to the persistence of low entropy gas that is unable
to cool. This gas by its very nature has not completely thermalised to the
global cluster temperature and has significant residual bulk kinetic energy. 
In the latter two runs, the non-gravitational heating generates a smoother
distribution that is evidently closer to hydrostatic equilibrium. 
%\vspace{-0.5cm}

\subsection{Effect of hydrostatic assumption on $Y_{500}$ and $Y_{\rm X,500}$}

\begin{figure}
\centering
\includegraphics[width=8.5cm]{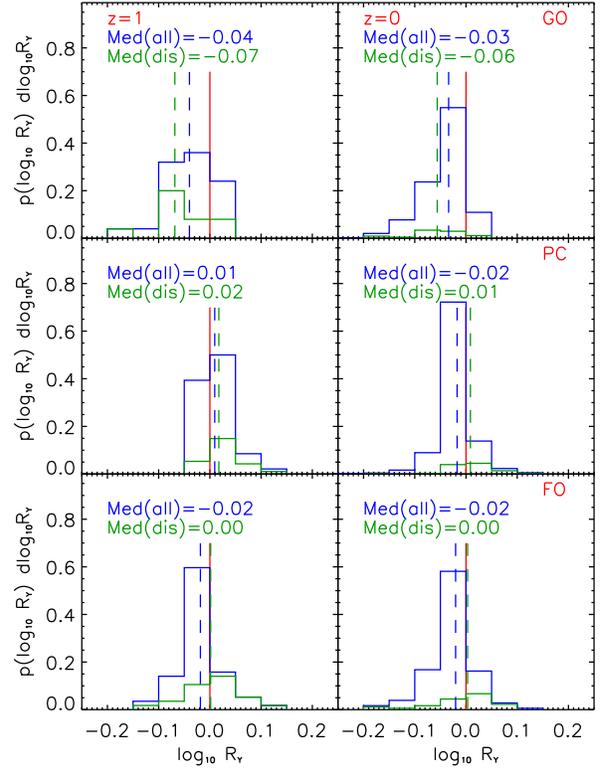}
\caption{As in Fig.~\ref{fig:mhseratio} but for the ratio of
estimated-to-true $Y_{500}$ values, $R_{\rm Y}$. The difference
in $\log(Y_{500})$ is very small in the runs with non-gravitational heating.}
\label{fig:yhseratio}
\end{figure}

The use of hydrostatic mass estimates also affects the SZ flux through
the use of $r_{500}^{\rm HSE}$ to define the cluster radius; a smaller
radius will result in a lower value for $Y$. We define a similar 
quantity to the mass ratio, $R_{\rm Y} \equiv 
\log_{10} (Y_{500}^{\rm HSE} / Y_{500})$, and present the distribution of
values in Fig.~\ref{fig:yhseratio}. Again, we present the ratio in this
way as it directly gives the shift in $\log_{10}Y$ values due to the hydrostatic
estimate.

As was the case with the total mass estimates, there is a larger bias
(and scatter)  in the $Y_{500}$ values for the GO run but the overall
effect is smaller as it is entirely due to the (small) shift in $r_{500}$.
The median $R_{\rm Y}$ is $-0.03$ for GO at $z=0$, 
increasing to only $-0.02$ for the PC and FO runs. Since $r_{500}^{\rm HSE}<r_{500}$
on average, the integrated flux is also smaller. Again, the results
are not significantly different at high redshift or when only the disturbed
clusters are selected. We have also checked the equivalent result for the $Y_{\rm X,500}$
values and they are very similar to the $Y_{500}$ results. 

\subsection{Estimated $Y_{500}-M_{500}$ relation directly from HSE}

\begin{figure}
\centering
\includegraphics[width=8.5cm]{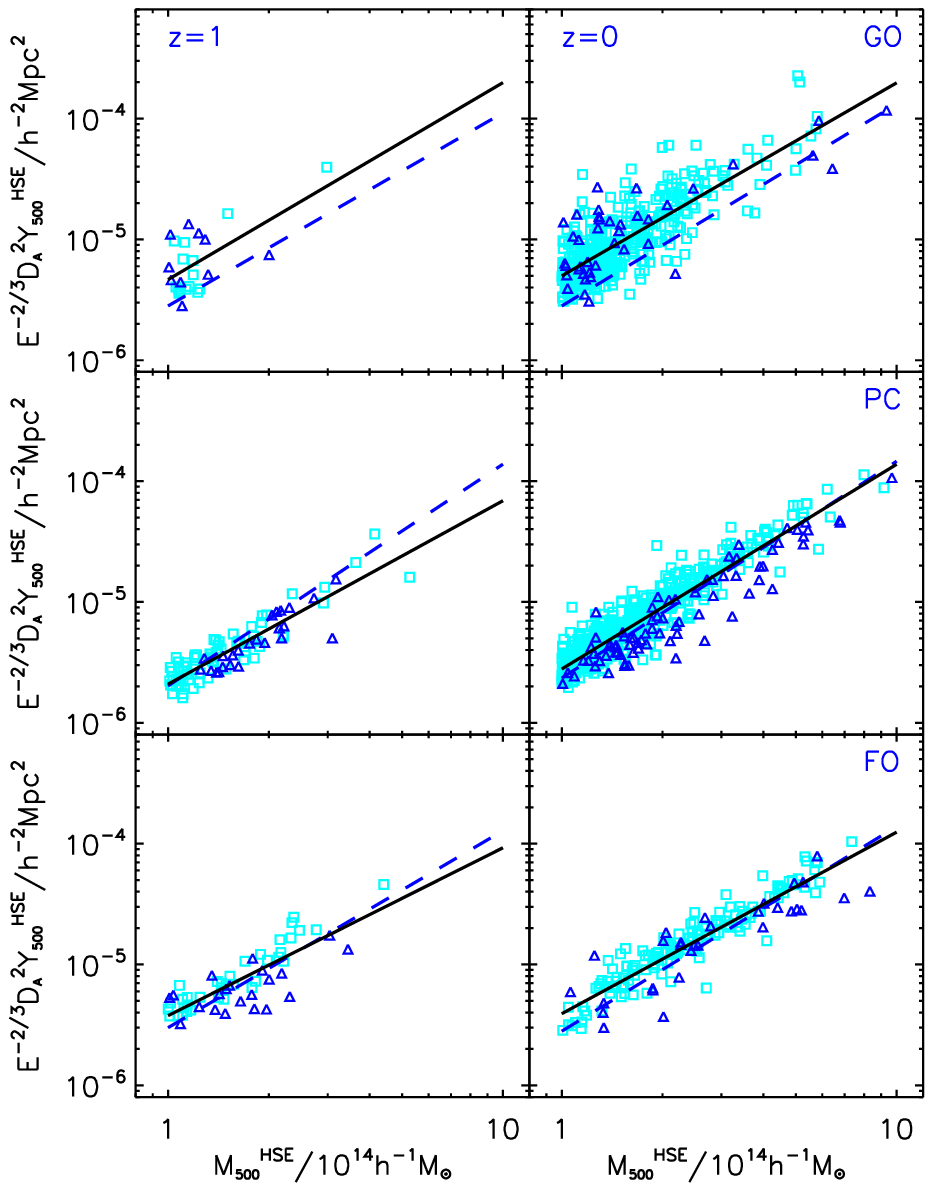}
\caption{The $Y_{500}^{\rm HSE}-M_{500}^{\rm HSE}$ relations (i.e. 
using estimated values for each cluster assuming the gas is hydrostatic) 
for the GO, PC and FO models at $z=1$ and $z=0$. 
The squares correspond to regular clusters while the triangles are disturbed clusters.
The solid line is the best-fit relation while the dashed line shows the best-fit
true $Y_{500}-M_{500}$ relation. Fits are performed for all clusters with
$10^{14}\hMsol<M_{500}^{\rm HSE}<10^{15}\hMsol$ and best-fit parameter values are
given in Table~\ref{tab:ymrelhse}. It is clear that the hydrostatic assumption 
is more robust for the PC and FO runs than for the GO run.}
\label{fig:y500m500hse}
\end{figure}

We now put together these results to study how the $Y_{500}-M_{500}$ 
relation is affected by the hydrostatic assumption. These results are shown in 
Fig.~\ref{fig:y500m500hse} for the GO, PC and FO runs at $z=1$ and $z=0$.
The best-fit $Y_{500}^{\rm HSE}-M_{500}^{\rm HSE}$ relation is shown as a 
solid line and we also plot the best-fit (true) $Y_{500}-M_{500}$ relation
as the dashed line in each panel. Values for the parameters describing 
the best-fit relations (normalisation, $A$; slope, $B$; and scatter, $\sigma_{\log_{10}Y}$)
are given in Table~\ref{tab:ymrelhse}.

The offset in $M_{500}$ values ($R_{\rm M}=-0.14$ at $z=0$) in the GO model is clearly visible in the
top-right panel of Fig.~\ref{fig:y500m500hse}, 
where the best-fit relation is offset to larger $Y_{500}$ values for
a given value of $M_{500}^{\rm HSE}$. The large spread in the $R_{\rm M}$ distribution is also
evident as the scatter has increased significantly ($\sigma_{\log_{10}Y}\simeq 0.19$, c.f. 
Fig.~\ref{fig:ymrelz10} where $\sigma_{\log_{10}Y}\simeq 0.04$). The offset is insensitive to mass
in this model, resulting  in a relation that has similar slope (1.6) to the true $Y_{500}-M_{500}$ relation.
The offset in normalisation has also led to a significant drop in the number of clusters in the sample
at each redshift; as a result there are only 98 clusters at $z=1$, making a reliable estimate of the
relation difficult (but the trends are nevertheless consistent with those seen at $z=0$).

The best-fit $Y_{500}^{\rm HSE}-M_{500}^{\rm HSE}$ relation from the PC run at $z=0$ is 
remarkably similar to the underlying relation, although the scatter has also increased considerably
to $\sigma_{\log_{10}Y} \simeq 0.11$. Results at $z=1$ prefer a flatter slope but this is 
somewhat affected by a few higher mass clusters (the slope is $1.5 \pm 0.1$). The estimated 
relation for the FO model is also similar to the true relation, with a preference for a slightly
flatter slope and larger scatter ($\sigma_{\log_{10}Y}\simeq 0.13$ at $z=0$).
The disturbed cluster sub-sample is most strongly biased in the PC results at $z=0$, where
the clusters have larger HSE masses for their flux, relative to the regular systems. 

\subsection{Estimated $Y_{500}-M_{500}$ relation using $Y_{\rm X,500}$}

When mass estimates are required for larger samples of clusters, the direct
hydrostatic method discussed above can be prohibitively expensive as it requires
the density and temperature profiles to be known out to $r_{500}$ and beyond.
An alternative, indirect method is to use a mass {\it proxy}, where mass
is estimated from a mass-observable scaling relation that is pre-calibrated 
using fewer clusters. Historically, $T_{\rm X}$ was the observable of choice but 
recent studies have focussed on the use of $Y_{\rm X}$ due to its low scatter
(\citealt{Kravtsov06}; see also \citealt{Arnaud07,Maughan07,Arnaud10,Sun11}). 
Indeed, the Planck Collaboration \citep{Planck11c} made use of the 
$M_{500}-Y_{\rm X,500}$ relation, calibrated by \citet{Arnaud10} from the REXCESS 
sample of 33 clusters, to estimate $r_{500}$ and $M_{500}$ for their larger 
(PXMM) sample of 62 clusters.

The procedure for estimating $M_{500}$ works as follows. Assuming that all clusters 
lie on an $M_{500}-Y_{\rm X,500}$ relation and that they evolve self-similarly with redshift, 
then $r_{500}$ may be found using
\begin{equation}
r_{500}(Y_{\rm X}) = 
\alpha \,
E(z)^{-4/5} \, m(Y_{\rm X})^{1/3} \hMpc,
\label{eqn:r500YX}
\end{equation}
where $\alpha$ is a known constant and 
$m(Y_{\rm X})$ is the best-fit value of $E^{2/5}M_{500}(Y_{\rm X,500})$ from the
scaling relation for a given $Y_{\rm X,500}$. The mass, $M_{500}(Y_{\rm X})$, can
then be estimated using equation~(\ref{eqn:m500hse}).
In practice, this equation must be solved iteratively: a value for $r_{500}$ is first guessed
then $Y_{\rm X,500}$ is calculated within this radius (from the integrated gas mass and
average X-ray temperature), allowing a new value for $r_{500}$ to be computed
from equation~(\ref{eqn:r500YX}). This is repeated until convergence is achieved. 
Since clusters do not all lie on this relation 
(even though this particular relation is chosen for its low scatter) the derived 
$r_{500}$ may be inaccurate for an individual cluster, but the overall relation 
should be unbiased.

\begin{figure}
\centering
\includegraphics[width=8.5cm]{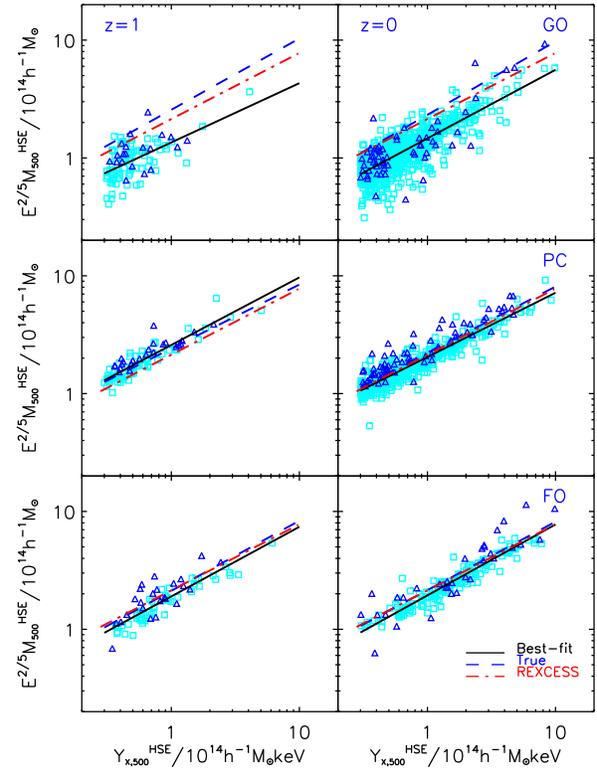}
\caption{As in Fig.~\ref{fig:y500m500hse} but for the 
$M_{500}^{\rm HSE}-Y_{X,500}^{\rm HSE}$ relation. Clusters are now
selected with $3\times 10^{13}<Y_{\rm X,500}^{\rm HSE}<10^{15}$ which,
for the PC and FO models, matches well to our normal mass range.
The solid line is the best-fit to the relation and the dashed
line the true relation. The dot-dashed line is the best-fit
relation from REXCESS ({\it XMM-Newton}) data as found by \citet{Arnaud10}.}
\label{fig:m500yx500hse}
\end{figure}

We have applied this procedure to our simulated clusters and will refer to
the resulting $Y_{500}-M_{500}$ relation as the
$Y_{500}^{Y_{\rm X}}-M_{500}^{Y_{\rm X}}$  relation. We first show our
derived $M_{500}^{\rm HSE}-Y_{\rm X,500}^{\rm HSE}$ relations, required
for equation~(\ref{eqn:r500YX}), for the three models at $z=1$ and $z=0$ 
in Fig.~\ref{fig:m500yx500hse}. Since $Y_{\rm X}$ (and not mass) is 
on the $x$-axis, we restrict our fits to clusters with 
$3\times 10^{13} \YXunits <Y_{\rm X,500}^{\rm HSE}< 10^{15} \YXunits$,
as this approximately matches our adopted mass range 
($10^{14}-10^{15}\hMsol$) for the PC and FO models.

Qualitatively, the same conclusions can be drawn as for the 
$Y_{500}^{\rm HSE}-M_{500}^{\rm HSE}$ relation 
(Fig.~\ref{fig:y500m500hse}): the GO model shows a large scatter and the relation
is offset due to the mass estimates being systematically low. As expected, however,
the PC and FO results agree very well with the best-fit underlying relation with
still relatively small scatter, $\sigma_{\log_{10}M}=0.06-0.08$. 
We also compare our results to the observed best-fit relation at low redshift 
from \citet{Arnaud10}, shown as the dot-dashed line. The agreement between the
PC and FO models is very good, with the latter prefering a slightly flatter slope
than the observed relation.

\begin{figure}
\centering
\includegraphics[width=8.5cm]{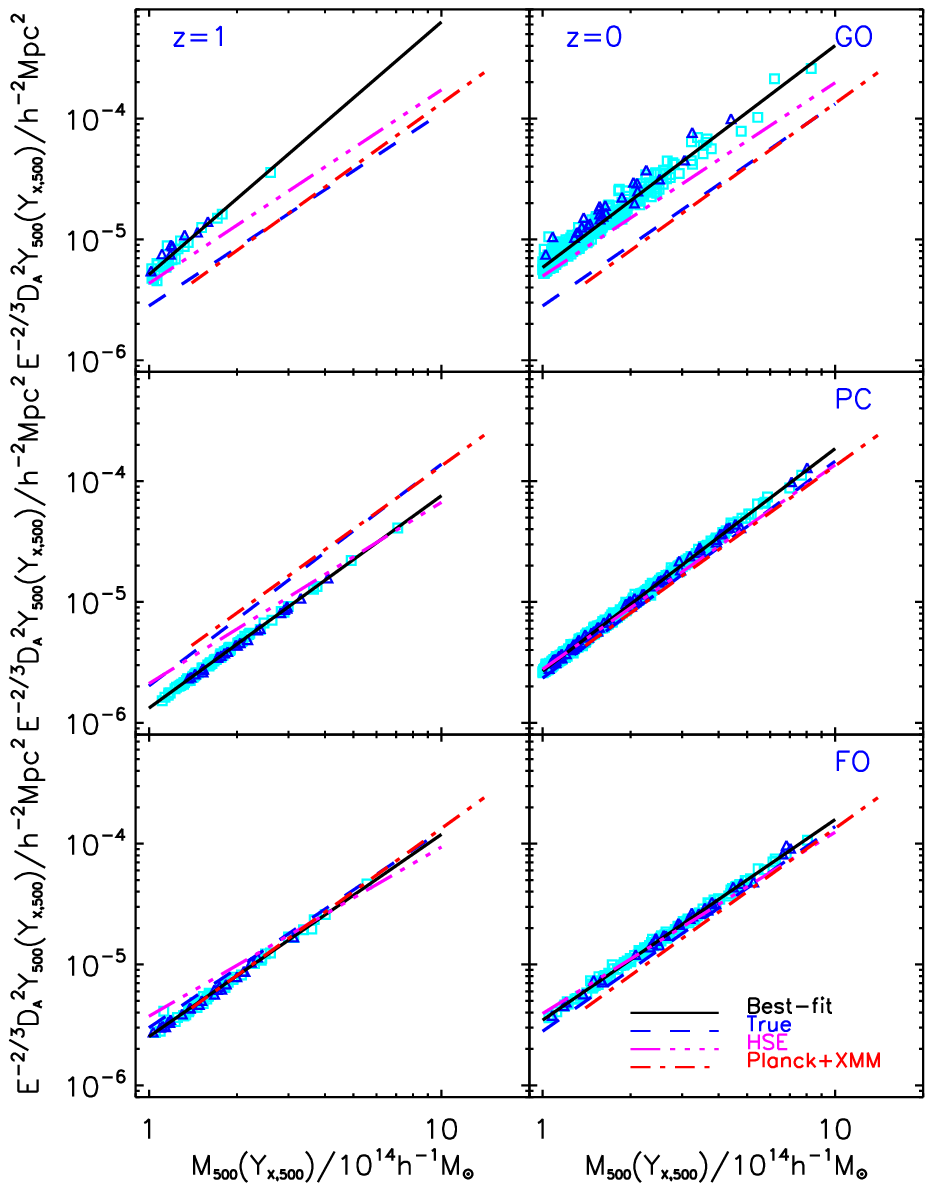}
\caption{$Y_{500}-M_{500}$ relations for the three models at $z=1$ and $z=0$
when $r_{500}/M_{500}$ is estimated from the best-fit $M_{500}^{\rm HSE}-Y_{\rm X,500}^{\rm HSE}$
relation with no scatter. We also show the best-fit true $Y_{500}-M_{500}$ relation (dashed line),
the best-fit $Y_{500}^{\rm HSE}-M_{500}^{\rm HSE}$ relation (triple-dot-dashed line) 
and the best-fit $z<0.5$ relation from PXMM data (\citealt{Planck11c}; dot-dashed line). 
The PC and FO models agree well with the observational data although there is a slight offset. The
scatter is much lower, however, due to the strong correlation between $Y_{500}$ and $Y_{\rm X,500}$.}
\label{fig:y500m500myx}
\end{figure}

The derived $Y_{500}^{Y_{\rm X}}-M_{500}^{Y_{\rm X}}$ relations are shown in
Fig.~\ref{fig:y500m500myx}. We have overlaid the best-fit relation to these
data (solid line); the best-fit true $Y_{500}-M_{500}$ relation (dashed line); 
the best-fit $Y_{500}^{\rm HSE}-M_{500}^{\rm HSE}$ relation (triple-dot-dashed line);
and the best-fit relation to the PXMM observational data (dot-dashed line). 

The final result is very striking for our PC and FO models. At $z=0$, there is only a small
amount of bias (around 20 per cent or so, or $\Delta \log_{10}Y_{500} \simeq 0.08$) with respect
to the underlying relation (and also, the PXMM relation). Such an offset, comparable to
the estimated intrinsic scatter in the observed relation, is small enough that it may 
just reflect our method not exactly matching that used by the Planck Collaboration.
Even  more striking is the reduction in scatter; $\sigma_{\log_{10}Y}\simeq 0.02$
for the PC and FO models at $z=0$, about half the size of the scatter in the true 
$Y_{500}-M_{500}$ relation. The reason for this reduction is obvious: from Fig.~\ref{fig:yyxrelz}
we saw that $Y_{500}$ and $Y_{\rm X,500}$ were strongly correlated, especially in the PC and 
FO models. Thus, a cluster that has a larger-than-average $Y_{\rm X,500}$ for its mass will
also have a larger-than-average $Y_{500}$. This was also true for the disturbed sub-sample
hence the reason why these clusters are unbiased with respect to the overall sample.
%\vspace{-0.5cm}

\section{Summary and Conclusions}
\label{sec:summary}

Large surveys are now being performed at millimetre wavelengths exploiting
the Sunyaev-Zel'dovich (SZ) effect to detect large samples of galaxy clusters out to high redshift.
Such samples will then be used to produce competitive constraints on cosmological parameters,
as well as to study the variation in physical properties of the intracluster gas (especially
the gas pressure) with mass and redshift. The cosmological application relies on the 
statistical estimation of cluster mass through the SZ $Y-M$ relation. 
In recent work (e.g. \citealt{Andersson11,Planck11c}) the first SZ-selected
samples of clusters have already been used to estimate the cluster $Y-M$ relation and
full results of cosmological analyses are expected over the next few years.

In this paper, we have analysed some of the largest $N$-body/hydrodynamic simulations of
structure formation (the Millennium Gas Simulations) to study the dependence of SZ cluster
properties on gas physics, at both low ($z=0$) and high ($z=1$) redshift. The large volume
used in these simulations produces significant (hundreds to thousands) samples of clusters 
over the interesting range of cluster masses ($10^{14}-10^{15} \hMsol$). We considered three
cluster gas physics models: a non-radiative (gravitational heating only) simulation that 
ought to produce an approximately self-similar cluster population; and two simulations that 
incorporate additional non-gravitational heating (a model that uniformly pre-heats the gas
at high redshift and a model that includes feedback
from stars and active galactic nuclei in galaxies). The feedback model is our most realistic, in that it has
already been shown to reproduce many of the scaling properties of X-ray clusters, especially
those with non-cool cores \citep{Short10}. 

We started by investigating the hot gas pressure profiles of our simulated clusters and
how they compare to the pressure profile advocated by \citet{Arnaud10}. We then compared 
our derived SZ scaling relations (between $Y_{500}$ and total mass, hot gas mass, X-ray temperature
and the X-ray analogue to the SZ $Y$ parameter, $Y_{\rm X}$) with the recent observational 
results, in particular those obtained from a combined SZ+X-ray analysis performed by the 
Planck Collaboration. We also tested two of the 
key assumptions used in the observed analysis, namely that the mean $Y_{500}-M_{500}$ relation is
unaffected by the assumption that the gas is hydrostatic and by the presence of any other
hot gas along the line-of-sight. Our main conclusions can be summarised as follows:

\begin{itemize}
\item In accord with previous studies, our 
simulation with non-radiative hydrodynamics produces a (spherical) $Y_{500}-M_{500}$
relation that has a self-similar slope (5/3) and also evolves with redshift according
to the self-similar expectation, $E(z)^{2/3}$. Simulations with non-gravitational 
heating (both pre-heating and feedback cases) create slightly steeper $Y_{500}-M_{500}$ relations 
(with a slope of $1.7-1.8$, when clusters across the mass range, 
$10^{14}\hMsol < M_{500} < 10^{15}\hMsol$, are considered) 
but the evolution with redshift is still close to self-similar. 

\item The simulations were compared with the {\it Planck+XMM} 
results at $z<0.5$ \citep{Planck11c} and very good agreement was found 
for a number of scaling relations  
($Y_{500}$ versus $M_{500}, M_{\rm gas,500}$ and $kT_{\rm X}$) for the pre-heating
and feedback models. The scatter in the $Y_{500}-M_{500}$ relation is smaller 
than observed, however, with $\sigma_{\log_{10}Y}\simeq 0.04$.

\item Intracluster gas in the non-radiative simulation contains a significant
unthermalised component, due to the presence of low-entropy, clumpy gas. This causes
an offset in the $Y_{500}-Y_{\rm X,500}$ relation, which tests the difference between
the mass-weighted and X-ray temperatures. As a result, hydrostatic mass estimates are
biased low by 20-30 per cent. The pre-heating and feedback simulations on the other hand
predict smoother gas distributions, with $Y_{500} \simeq Y_{\rm X,500}$ and much smaller
hydrostatic bias (estimated masses are only $\sim 10$ per cent lower).

\item The estimated $M_{500}-Y_{\rm X,500}$ relations (assuming the gas is hydrostatic)
are in good agreement with the recent observational determination by \citet{Arnaud10}.
When $Y_{\rm X,500}$ is used as a mass proxy to predict the SZ $Y_{500}-M_{500}$ relation,
only a small ($\sim 20$ per cent) offset in normalisation
from the true relation (and thus the observed relation from 
{\it Planck+XMM} data) is found. The scatter in the recovered relation is very small
($\sigma_{\log_{10}Y}\simeq 0.02$) due to the strong correlation between $Y_{500}$ and 
$Y_{\rm X,500}$. Clusters that are undergoing major mergers are not significantly 
offset from the mean relation.

\item Hot gas pressure profiles are well described by generalised NFW profiles, as
suggested by \citet{Nagai07b} and show that the majority of the contribution to the
SZ $Y$ parameter (where $r^3 P(r)$ is maximal) comes from radii close to $r_{500}$.
Splitting the cluster samples into low and high-mass
sub-samples, we find little difference between the two in the run with non-radiative
hydrodynamics, as expected. The runs with non-gravitational heating predict that
low-mass clusters have lower core pressures and higher pressures in the cluster outskirts, 
when scaled according to the self-similar expectation. This non-self-similar-behaviour
can be attributed to the heating that is more effective in low-mass clusters and 
acts to push the gas out to large radii. There is also significant cluster-cluster
scatter, especially in the core region and in the outskirts, where individiual pressure
profiles can be 50 per cent higher than the median profile. 

\item We also compared our median pressure profiles
with the Arnaud et al. profile and found good agreement (within 10 per cent) for our high-mass
clusters at $r>0.5r_{500}$, in the pre-heating and feedback models. Low-mass clusters
are especially discrepant in the core regions, likely due to the absence of radiative cooling
in our models. Using the X-ray temperature (rather than hot gas mass-weighted temperature) in
the pressure calculation, as well as using hydrostatic estimates of $r_{500}$ and $M_{500}$,
only makes a significant ($>10$ per cent) difference to the non-radiative simulation for
the reasons already mentioned.

\item Finally, we considered the effects of projection due to large-scale structure
along the line-of-sight, by analysing 50 $5^{\circ} \times 5^{\circ}$ maps of the thermal 
SZ effect. By measuring the cylindrical SZ flux associated with each cluster and comparing
to the flux from the cluster region alone, we were able to discern the contribution from
additional structures, in the non-radiative and pre-heating simulations. The pre-heating
model showed the largest bias, where low-mass clusters ($M_{500}\simeq 10^{14}\hMsol$)
had cylindrical $Y_{500}$ values that were around 2-3 times higher than the value
from the cluster region. This is due to the large amount of thermal energy injected
into the gas at high redshift, as evidenced by the three-fold increase in the mean-$y$
parameter. Subtracting the contribution from an assumed mean background we find the
recovered $Y_{500}-M_{500}$ relation to be unbiased with respect to the cluster relation,
with some additional scatter that is model-dependent.
\end{itemize}

In summary, we can conclude that when our more realistic models for the intracluster gas 
are employed (namely those that raise the entropy of the gas to match global X-ray 
scaling relations), the SZ $Y-M$ relation is in good agreement with the observations (Fig.~\ref{fig:ymrelz10})
and is largely unaffected by two of the main sources of systematic uncertainty: hydrostatic
bias (Figs.~\ref{fig:y500m500hse} and \ref{fig:y500m500myx}) and projection effects from large-scale structure 
(Fig.~\ref{fig:y500m500cylbsub}).

While our analysis has been one of the most comprehensive to date and used some of the
largest and most sophisticated simulation models, there are some significant 
shortcomings that still need to be addressed. Firstly, the effects of radiative cooling 
were not included in our most realistic (feedback) model, so the model cannot yet 
match the full X-ray cluster population (namely the brightest objects with cool cores,
\citealt{Short10}). As we argued, this omission is likely not a significant problem 
for the $Y-M$ relation but will affect the hot gas pressure profile so it should be 
addressed in future work. Secondly, we were unable to test 
projection effects for the feedback model as we only have a sample of clusters rather
than the full cosmological volume. Finally, the cosmological model adopted for the 
simulations (identical to that used in the original Millennium Simulation)
is no longer favoured; in particular the value of $\sigma_8$ is higher than the current
best estimate ($\sigma_8=0.9$ in the simulations, c.f. $\sigma_8\simeq 0.8$ from the 
{\it WMAP} 7-year data; \citealt{Komatsu11}). 
Using the presently-favoured cosmological model is likely to 
reduce the scale of projection effects, however, as in it structure formation will be less advanced.

We are currently preparing a new generation of Millennium Gas simulations that will 
rectify all of these problems, starting with a new version of our existing feedback 
model that will deal with the second and third issues. This new simulation, which 
is also being run at higher resolution and with an updated semi-analytic galaxy 
formation model \citep{Guo10}, will additionally allow cosmologically-dependent 
statistical predictions for the SZ signal to be performed, namely the
SZ power spectrum.
%\vspace{-0.5cm}

\section*{Acknowledgements}
We thank the reviewer, Stefano Borgani, for his insightful comments.
The simulations used in this paper were performed at the University of Nottingham 
HPC Facility and on the ICC Cosmology Machine, which is part of the DiRAC Facility jointly
funded by STFC, Large Facilities Capital Fund of BIS and Durham University.
STK, MWP and RAB were supported by STFC through grant ST/G002592/1.
CJS, PAT and ARL was supported by STFC through grants 
ST/F002858/1 and ST/I000976/1. OEY was supported by an STFC quota studentship.
%\vspace{-0.5cm}

\appendix

\section{Evolution of scaling relations}

\begin{figure*}
\centering
\includegraphics[width=14cm]{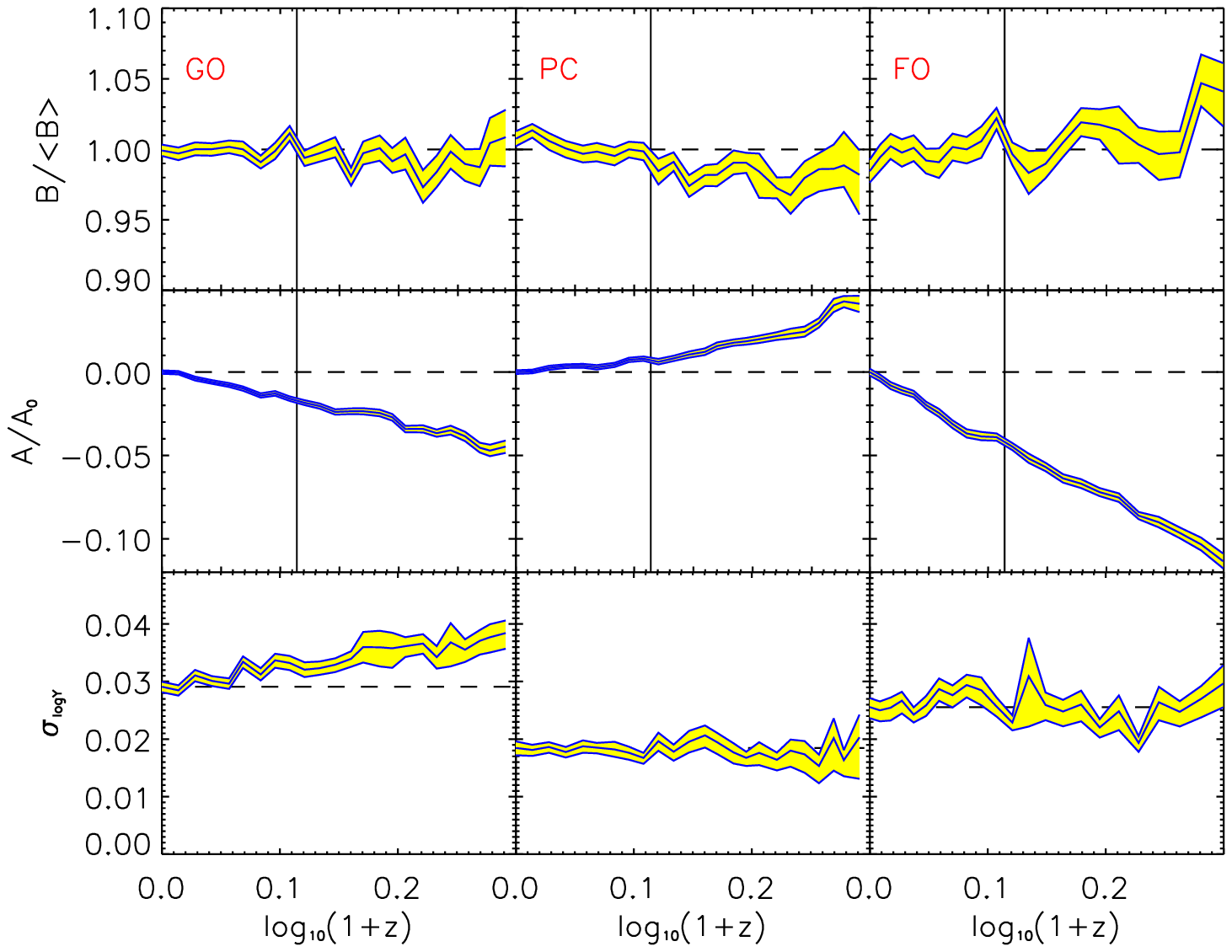}
\caption{As in Fig.~\ref{fig:ymrel_evol} but for the $Y_{500}-M_{\rm gas,500}$ relation.}
\label{fig:ymgasrel_evol}
\end{figure*}

\begin{figure*}
\centering
\includegraphics[width=14cm]{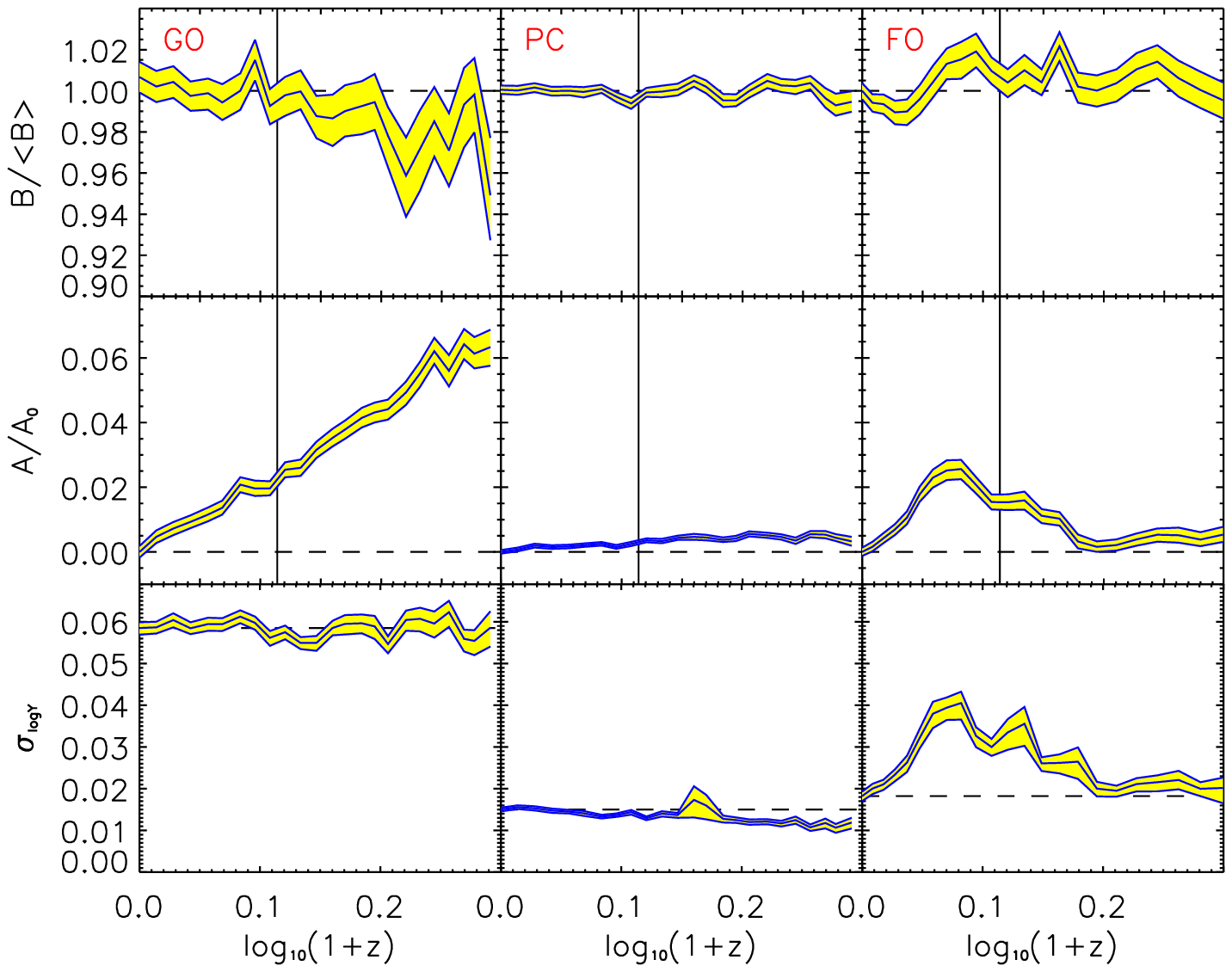}
\caption{As in Fig.~\ref{fig:ymrel_evol} but for the $Y_{500}-Y_{\rm X,500}$ relation.}
\label{fig:yyxrel_evol}
\end{figure*}

The following figures (Figs.~\ref{fig:ymgasrel_evol},~\ref{fig:ytrel_evol} and
\ref{fig:yyxrel_evol}) illustrate the evolution of the slope, normalisation and
scatter with redshift for the $Y_{500}-M_{\rm gas,500}$, $Y_{500}-T_{\rm sl}$
and $Y_{500}-Y_{\rm X,500}$ relations respectively. Details of what is plotted
in each panel are identical to Fig.~\ref{fig:ymrel_evol} and are discussed
in Section~\ref{sec:yobs}.

\begin{figure*}
\centering
\includegraphics[width=14cm]{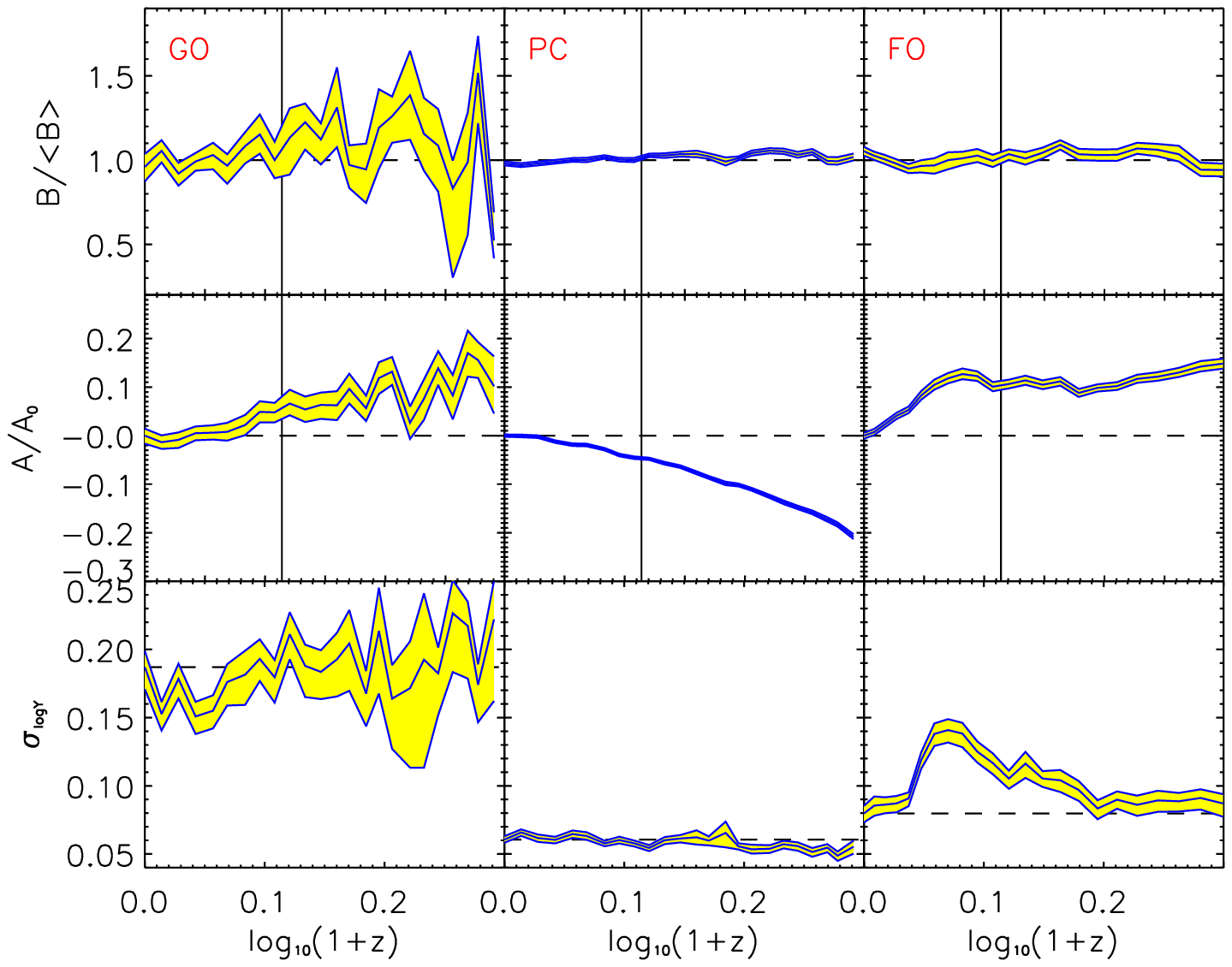}
\caption{As in Fig.~\ref{fig:ymrel_evol} but for the $Y_{500}-T_{\rm sl}$ relation.}
\label{fig:ytrel_evol}
\end{figure*}

\bsp

\label{lastpage}
\end{document}